\documentclass[fleqn,usenatbib]{mnras}

\usepackage[T1]{fontenc}

\DeclareRobustCommand{\VAN}[3]{#2}
\let\VANthebibliography\thebibliography
\def\thebibliography{\DeclareRobustCommand{\VAN}[3]{##3}\VANthebibliography}

\usepackage{graphicx}	% Including figure files
\usepackage{amsmath}	% Advanced maths commands
\usepackage{amssymb}	% Extra maths symbols
\usepackage{xcolor}
\usepackage{rotating}
\usepackage{multirow}
\usepackage{newtxtext,newtxmath}
\usepackage{tablefootnote}
\newcommand{\teff}{$T_{\mathrm{eff}}$}
\newcommand{\muhz}{$\mu$Hz}
\newcommand{\numax}{$\nu_{\mathrm{max}}$}

\newcommand{\dnu}{$\Delta\nu$}

\newcommand{\msol}{M$_\odot$}

\newcommand{\kepler}{\textit{Kepler}}
\newcommand{\ktg}{K2 GAP}

\title[Fast \dnu{} Vetting with Neural Networks]{Vetting Asteroseismic \dnu{} Measurements using Neural Networks}
\author[C. Reyes et al.]{
Claudia Reyes$^{1}$\thanks{E-mail: claudiarreyes@icloud.com}, 
Dennis Stello$^{1,2,3}$, 
Marc Hon$^{4,1}$ and
Joel C. Zinn$^{5,1}$\thanks{NSF Astronomy and Astrophysics Postdoctoral Fellow}\\
$^{1}$School of Physics, University of New South Wales, NSW 2052, Australia\\
$^{2}$Sydney Institute for Astronomy (SIfA), School of Physics, University of Sydney, NSW 2006, Australia\\
$^{3}$Stellar Astrophysics Centre, Department of Physics and Astronomy, Aarhus University, DK-8000 Aarhus C, Denmark\\
$^{4}$Institute for Astronomy, University of Hawaii, 2680 Woodlawn Drive, Honolulu, HI 96822, USA\\
$^{5}$Department of Astrophysics, American Museum of Natural History, Central Park West at 79th Street, New York, NY 10024, USA\\
}
\date{Accepted XXX. Received YYY; in original form ZZZ}

\pubyear{2021}

\begin{document}
\label{firstpage}
\pagerange{\pageref{firstpage}--\pageref{lastpage}}
\maketitle

\begin{abstract}
Precise asteroseismic parameters allow one to quickly estimate radius and mass distributions for large samples of stars.
A number of automated methods are available to calculate the frequency of maximum acoustic power (\numax) and the frequency separation between overtone modes (\dnu) from the power spectra of red giants. 
However, filtering through the results requires either manual vetting, elaborate averaging across multiple methods, or sharp cuts in certain parameters to ensure robust samples of stars free of outliers. 
Given the importance of ensemble studies for Galactic archaeology and the surge in data availability, faster methods for obtaining reliable asteroseismic parameters are desirable. 
We present a neural network classifier that vets \dnu\ by combining multiple features from the visual \dnu\ vetting process. Our classifier is able to analyse large numbers of stars determining whether their measured \dnu\ are reliable thus delivering clean samples of oscillating stars with minimal effort. Our classifier is independent of the method used to obtain \numax\ and \dnu, and therefore can be applied as a final step to any such method.
Tests of our classifier's performance on manually vetted \dnu\ measurements
reach an accuracy of 95\%. 
We apply the method to giants observed by K2 Galactic Archaeology Program and find that our results retain stars with astrophysical oscillation parameters consistent with the parameter distributions already defined by well-characterised \kepler\ red giants.

\end{abstract}


%%%%%%%%%%%%%%%%%%%%%%%%%%%%%%%%%%%%%%%%%%%%%%%%%%

%%%%%%%%%%%%%%%%% BODY OF PAPER %%%%%%%%%%%%%%%%%%

\section{Introduction}
Since the launch of  CoRoT 
\citep{2000JApA...21..319B} %\citep{2008Sci...322..558M, 2009Natur.459..398D} 
and \kepler{}  \citep{2010Sci...327..977B, 2010ApJ...713L..79K}, asteroseismic analysis pipelines such as SYD \citep{2009CoAst.160...74H}, COR \citep{2009A&A...508..877M}, CAN \citep{2010A&A...522A...1K}, A2Z \citep{2010A&A...511A..46M,2014A&A...568A..10G}, BAM \citep{2019ApJ...884..107Z}, and BHM \citep{ 2017MNRAS.466.3344E} have been developed to extract \numax\ and \dnu\ in more automated ways than was done in the past. 
To analyse the data and to determine \dnu, each of these pipelines relies on different methods such as: %the autocorrelation of the power spectrum (SYD, BAM), the autocorrelation of the timeseries (COR), a fit to the power spectrum (CAN), a fit to the folded power spectrum (BAM), and the power spectrum of the power spectrum (A2Z, BHM). Additional statistical testing is also used by some as internal detection method}.
the autocorrelation of the power spectrum (SYD), the autocorrelation of the timeseries (COR) or equivalently, the power spectrum of the power spectrum (BHM), a fit to the power spectrum (CAN), a fit to the folded power spectrum (BAM), or a combination thereof (A2Z, BAM). Additional statistical testing is also used by some as internal detection calibration. 
%\textbf{Each of these pipelines relies on different methods to analyse the data. For determining \dnu\ SYD uses the autocorrelation of the power spectrum; COR, the autocorrelation of the timeseries; CAN uses a fit to the power spectrum; A2Z, the power spectrum of the power spectrum; BAM, a fit to the folded power spectrum and the autocorrelation of the power spectrum; and BHM uses the power spectrum of the power spectrum. }
%\textbf{COR, CAN and BAM are also internally calibrated by means of statistical testing.}
However, many of the pipelines still require a form of vetting to remove unreliable measurements of \dnu\ beyond what is captured by statistical significance %\textbf{detection} 
testing. The vetting therefore often involves some sort of manual verification sometimes involving results from multiple pipelines and/or hard-coded cuts in certain parameters. The former is very time consuming and the latter can easily result in unphysical sharp features in the properties of the resulting stellar population, which can be undesirable.

After \kepler{} came K2 \citep{2014PASP..126..398H}, and for the first time a constant flow of large amounts of data of previously unknown seismic targets that needed vetting beyond what is suitably performed using a manual approach. 
With the launch of TESS \citep{2014SPIE.9143E..20R} and later in this decade of PLATO \citep{2017EGUGA..19.4829R}, fully automated yet robust methods are more necessary than ever to ensure fast and reliable asteroseismic measurements providing both complete and pure sets of measurements.

Here we present a neural network-based classifier that is able to determine whether \dnu{} values are reliable, independent of the method used to derive \dnu{} and \numax{}. 
We start by giving an overview of the data used in this paper. Then we describe the methods used to build the machine learning model, and next we show its performance on the training set using traditional machine learning performance metrics. Finally, we examine the classifier's performance on data from different pipelines by comparing our vetted results with \numax{} and \dnu{} distributions reported by \citet{zinn2021k2}.

\section{Data}
\label{sec:data}
The data used in this project correspond to observations obtained as part of the K2 Galactic Archaeology Program, GAP, Campaigns 1 to 8 and 10 to 18 \citep{2017ApJ...835...83S, 2020ApJS..251...23Z, zinn2021k2}. 
K2 GAP targets were chosen to satisfy simple colour and magnitude cuts where red giants are more likely to be found \citep{2015ApJ...809L...3S, 2021arXiv210912173S}. Reasons for targeting red giants are: their high luminosity, which allows probing deeper Galactic regions, and their oscillation frequencies, which are detectable from K2 long-cadence data (which has a Nyquist frequency $\approx$ 280 \muhz{}).
Hence solar like oscillators from \ktg{} are expected to be mostly red giant branch (RGB) stars in various evolutionary phases as well as core Helium burning red clump (RC) stars.

For the initial analysis of this paper, values of \dnu{} and \numax{} are from the SYD pipeline, while in Sections \ref{sec:pipelines} we examine the results of our method on results from various other pipelines in literature.

\section{Methodology}
\label{sec:method}

We want to build a classifier that 
%is pipeline-agnostic, meaning that it 
can vet \dnu\ regardless of the pipeline that provided the measurement. %\textbf{ mimicking the visual vetting process}
For this task we choose to use supervised learning, a machine learning technique characterised by its use of labelled data sets to train algorithms for data classification (or regression). Specifically, we use neural networks because we need a method that can deal with the various aspects and complexities shown by oscillation spectra of red giants. Key capabilities of such non-linear algorithms, including parallel processing of multiple features and complex-pattern detection and extrapolation, make neural networks well-suited for this task.
%%\textbf{The visual analysis of a star's oscillation spectrum constitutes a traditional way of verifying the reliability of \dnu\ results.
%While there are limitations inherent to this method, such as difficult detections due to the proximity to the Nyquist frequency of the oscillation spectra of less evolved red giants and by the effects of low frequency resolution in the oscillation zone of the most evolved red giants, it remains a trusted and preferred tool by many, particularly in cases where other methods return substandard results.}

%\textbf{A big drawback is that visual vetting can be highly time consuming, however machine learning can help solve this problem: implementations of such algorithms learn from the data they are trained on and create a simplified version of the observations which are likely to generalise to new instances. 
%In particular deep neural networks are well-suited for this task thanks to their non-linearity and standout capabilities that include parallel processing of multiple features and complex-pattern detection and extrapolation.
%A classifier trained this way will be able to efficiently mimic visual vetting of \dnu\ regardless of the pipeline that provided the measurement.}

We first discuss the data set used to train our machine learning method in Section \ref{sec:training}, then we discuss the features we extract from power spectra in Section \ref{sec:inputs}, the neural network architecture in Section \ref{sec:model_arch}, the training steps in Section \ref{sec:train} and finally the assessment of the network's performance on the training data in Section \ref{sec:perf}.

\subsection{Training Set Preparation}
\label{sec:training}

\begin{figure*}
\includegraphics[width=\textwidth]{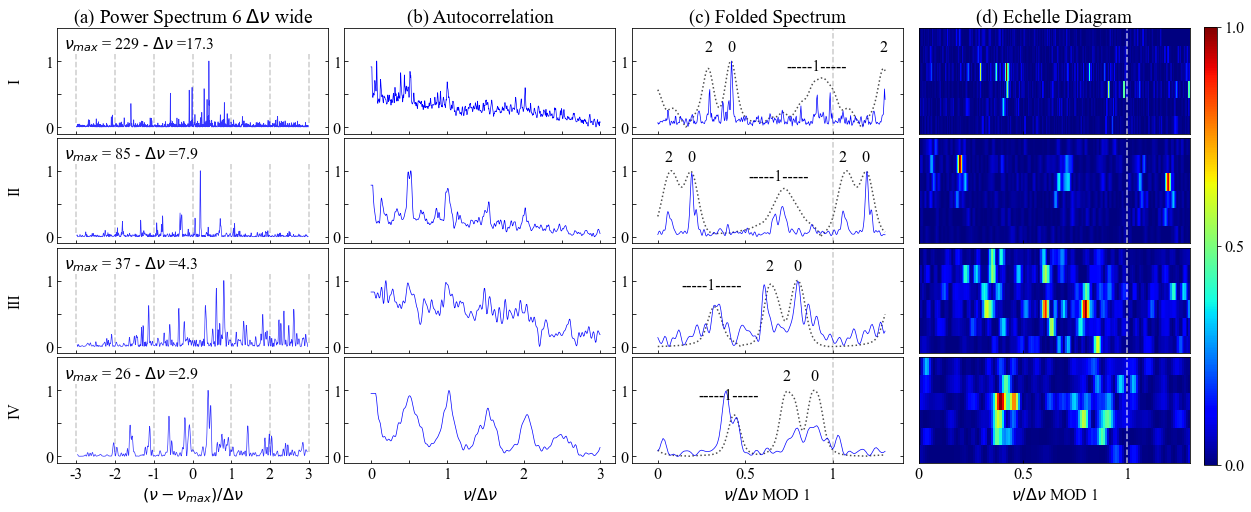}
    \caption{
    Examples of diagnostic plots from our training set showing stars with reliable \dnu\ in different \numax\ ranges. Rows marked I-IV correspond to stars from \ktg{} campaign 1: EPIC 201829369, 201207669, 201245474, 201681005 respectively. (a) portion of the power spectrum around the frequency of maximum power \numax{}. Segmented lines indicate multiples of \dnu. Here \numax\ has been shifted to the nearest multiple of \dnu. Annotated \numax{} and \dnu{} values are given in \muhz{}.(b) autocorrelation function showing the periodic peaks at multiples of \dnu{}, (c) folded spectrum of width 1\dnu{} obtained from folding the six central \dnu\ segments from (a), shown here together with the folded modelled spectrum (grey dotted outline) assigned to the star based on its \dnu. Modes $l=0,1,2$ are annotated. Modes of degree $l=1$ appear spread out showing the coupling of dipoles with a large number of higher order g modes from the core. (d) The échelle diagram constructed from the power spectrum from column (a), with hotter colours indicating higher relative power. The extension to the right of the grey line marking 1\numax\ on (c) and (d) mirrors the first 30\% of the diagram, for continuity. All data are scaled in amplitude between 0 and 1. }
    \label{fig:sampler}
\end{figure*}
%\textbf{Any bias introduced during training can be expected to propagate to our results, therefore it}
It is essential to carefully build a training set that is balanced and representative of the different spectra that our machine learning algorithm will encounter in practice.
Following \citet{2018ApJS..236...42Y} we use visual inspection to classify stars as having \dnu\ detections or not. %visual inspection to label and construct our training \textbf{sample.} %because power spectra are complex and there is no one established vetting method that does as well as the human eye. For instance, the sample of 16,000 \kepler\ red giants by \citet{2018ApJS..236...42Y}, relied on visual vetting as well. 
To prepare for the visual vetting we generate diagnostic plots that allow us to examine the identified oscillations in power spectra using the following three diagrams.

\subsubsection{Autocorrelation function}
\label{ssec:acf}
The first diagnostic we use is the autocorrelation function
(Figure\ref{fig:sampler}b) of the power spectrum (Figure\ref{fig:sampler}a). The autocorrelation highlights the near regularity of the oscillation spectrum and is useful to confirm if \dnu{} can be measured reliably and whether it has been measured correctly. This regularity is expected from the asymptotic relation \citep{1980ApJS...43..469T}, where the frequency of a mode with spherical degree $l$ of radial order $n$ is given by

    \begin{equation}
        \nu_{n,l} \approx \Delta \nu (n +\frac{l}{2} + \epsilon) - \delta \nu_{0,l}.
    	\label{eq:asymp}
    \end{equation}
\noindent Here $\epsilon$ is a dimensionless offset or phase term and the small separation $\delta \nu_{0,l}$ is defined as 0 for $l=0$. 
From equation \ref{eq:asymp} we expect the autocorrelation to show peaks
at multiples of $\sim$\dnu/2.

To produce the plots we calculate the autocorrelation up to a shift of the spectrum equivalent to three \dnu\ and then we scale the amplitudes between 0 and 1, but first we make sure to avoid the global maximum of the autocorrelation (at a shift of zero) by ignoring the first 0.02\dnu\ shift. To do this, we standardise the function to a fixed length ensuring that the autocorrelation peaks were not smoothed away. A length of 300 data points was found to be conservative, therefore the first 6 points are ignored for each star. %. Then, the first six points of the autocorrelation function are set to the value of the seventh point. However, for stars with \dnu\ smaller than 7\muhz, we set the first three data points to the value of the fourth because for such low \dnu\ stars the oscillation power is distributed over fewer frequency bins.

\subsubsection{Folded Spectrum}
\label{sec:FS}
The second diagnostic is the folded oscillation spectrum
(Figure\ref{fig:sampler}c). %It is constructed by taking the central 6\dnu-wide segment of the spectrum around the frequency of maximum power and co-adding each \dnu\ segment. 
It is constructed by taking the central portion of the spectrum around the frequency of maximum power and co-adding each \dnu\ segment. 
This provides a simple way of showing the regularity in the mode pattern, and is particularly useful for low S/N cases where not every segment on their own would necessarily show the full pattern of modes.

To further guide the eye towards the expected pattern, we use model spectra that we fold and lay on top of the observed folded spectra (see grey dotted lines in Figure\ref{fig:sampler}c). If a reliable measurement of \dnu\ is used, the overall shape described by the folded spectrum will follow this template. 
We used theoretical oscillation modes for 1\msol\ models in different evolutionary phases from the base to the tip of the RGB taken from \citet{2014ApJ...788L..10S}. These models were based on simulations from the ASTEC stellar evolution code \citep{2008Ap&SS.316...13C}, which does not include the later core Helium burning phases.

\begin{figure}
\includegraphics[width=\columnwidth]{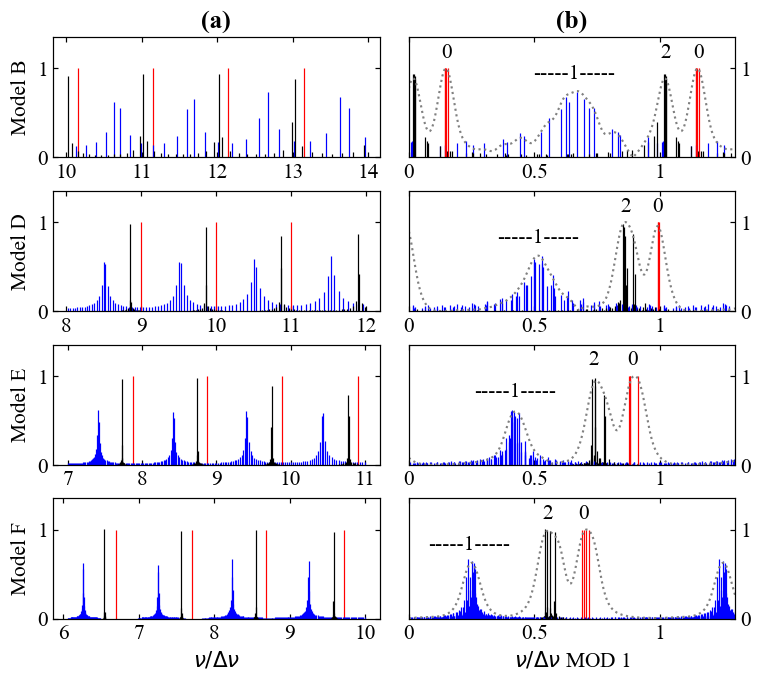}
    \caption{
    %THERE IS NO INDICATION OF THE Y-AXIS UNITS
    Models B, D, E and F from Table \ref{tab:table_models} representing four of the most commonly used models. Radial modes are shown in red, dipole modes in blue and quadrupole modes in black. All amplitudes are normalised.
    (a) four central segments of the spectra of width 1\dnu\ around the frequency of maximum oscillation power \numax. (b) %X-AXIS UNIT LOOKS STRANGE
    folded model spectra obtained from folding (a). The positions of the $l=0,1,2$ modes are annotated. The ordinate range of the folded spectrum is extended beyond 1 for clarity of the repeating structure of the oscillations.}
    \label{fig:tfs2}
\end{figure}

Table \ref{tab:table_models} lists the parameters of the models, from model A (\numax=260.5\muhz) to model I (\numax=4.0\muhz)
used to cover the entire range of frequencies in the K2 sample. We derived \dnu\ from the radial modes following the approach by \citet{2011ApJ...743..161W}, 
performing a weighted linear fit to the radial frequencies $l=0$ as a function of the order $n$ as in equation \ref{eq:asymp}, with weights obtained from a Gaussian window of width=0.25\numax\ centred on \numax, taking the slope of this fit as our \dnu.
The last column of Table \ref{tab:table_models} refers to the range of \dnu\ from real stars for which we use each model. %The limits in \dnu\
The boundaries for each range are set to the midpoint between \dnu\ of the models on a logarithmic scale.
Note that these models fully take into account the presence of mixed modes that arise from the coupling between pressure and gravity waves, as it is evident specially for $l=1$ from Figure \ref{fig:tfs2}a, in blue.
The peak height of each mode is modelled as the inverse of the square root of the mode's inertia and scaled to that of the radial modes interpolated at each mode frequency, following \citet{2010aste.book.....A} and \citet{2014ApJ...788L..10S}.

\begin{table}
    \caption{Nominal \numax{} values and calculated \dnu{} for the nine theoretical oscillation models considered to span the entire frequency range of the K2 sample. The last column shows the range of the observed \dnu{} we used to select which model corresponds to each star.}
	\centering
	\label{tab:table_models}
	\begin{tabular}{lccccr} 
		\hline
		Model & \numax\ [\muhz] & $\Delta\nu_{\mathrm{model}}$ [\muhz] & $\Delta\nu_{\mathrm{obs}}$ range  [\muhz] \\
		\hline
		 A & 350.5 & 24.5 & [ 18.9, 31.5 )\\
		 B & 180.8 & 14.7 & [ 11.4, 18.9 )\\
		 C & 94.6 & 8.9 & [ 7.65, 11.4 )\\
		 D & 64.2 & 6.6 & [ 5.35, 7.65 )\\
		 E & 37.4 & 4.4 & [ 3.44, 5.35 )\\
		 F & 20.2 & 2.7 & [ 2.08, 3.44 )\\
		 G & 10.0 & 1.6 & [ 1.25, 2.08 )\\
		 H & 5.5 & 1.0 & [ 0.88, 1.25 ) \\
		 I & 4.0 & 0.8 & [ 0.71, 0.88 ) \\
		\hline
	\end{tabular}
\end{table}

Figure \ref{fig:tfs2}a shows the central four \dnu\ segments around \numax\ of the modelled oscillation spectrum for models B, D, E and F, which represent four of the most commonly used models. 
%From these four central segments we obtain the folded spectra in Figure \ref{fig:tfs2}b. Although we use six segments around \numax\ for the real data, for low \numax\ models there are only significant oscillation modes in the four central \dnu\ segments around \numax. We therefore choose to consider only four \dnu\ segments.
Because for very low \numax\ models there are only significant oscillation modes in the four central \dnu\ segments around \numax\ (\citealt{2014ApJ...788L..10S}, Figure 1), we obtain the folded spectra in Figure \ref{fig:tfs2}b from these four segments. For the real data, however, we consider six central segments to account for %one-\dnu\ offsets of the power in either direction with respect to \numax\ that might be caused by an 
possible asymmetric distributions of the oscillation power around the value of \numax.% In practice, increasing the number of segments from 4 to 6 increased the ability of features based on the folded spectrum alone (Section \ref{sssec:inputsXC}) to discriminate reliable \dnu\ from tests on the training set by a few points of a percent.}
%, which is explained and expected due to the stochastic nature of the oscillations}

To obtain %a smooth template 
the templates corresponding to the dotted lines in Figure \ref{fig:tfs2}b, we generate an array (not shown) in which we represent each of the modes with a Lorentzian of width 0.04\dnu. %, where each point in the final array is the maximum of all the Lorentzians at that point (the upper envelope). 
Finally, we convolve %this upper envelope array 
the upper envelope of this array with a Gaussian of $\sigma = 7$ 
%\textbf{between 6 and 11} 
$\nu/\Delta\nu$ modulo 1, chosen to produce a smoothed template without losing any of the general features of the folded models.
The resulting %smooth 
template is scaled to a fixed range between 0 and 1 and used together with the star's folded spectrum as in Figure \ref{fig:sampler}c, where the template has been shifted to the position of maximum correlation. 
This shift takes care of the %unknown 
difference in $\epsilon$ between the data and the model \citep{2011ApJ...743..161W}. The choice of which template model to use is made according to each star's \dnu\ and Table \ref{tab:table_models}.
Because solar-like oscillations are stochastically driven, the oscillation amplitudes seen in short time series (such as the 80-day K2 data used in this work) can show significant variation from the simple, and rather regular, mode inertia-based `amplitudes' we use in the template. 
Hence we allow for some variation of the mode heights around the predicted model when we decide whether \dnu\ is reliable.

\subsubsection{Échelle diagram} 
\label{sec:echelle313}
The third diagnostic plot is the échelle diagram, which is created by dividing the power spectrum into segments of length \dnu{} and stacking each segment above one another. The resulting two-dimensional array is colour-coded according to power in each array bin. If a reliable measurement of \dnu\ is used, the $l=0$ and $l=2$ modes are expected to align vertically in the échelle diagram \citep{bedding2011solarlike}. Figure \ref{fig:sampler}d shows the diagram created from the central 6 \dnu\ segments of the power spectrum. 
The échelle diagram carries additional information that the autocorrelation and folded spectra do not provide; in particular, 
while the natural curvature of the mode pattern can slightly blur the autocorrelation function and the folded spectrum, it can be displayed neatly in the échelle diagram (\citealt{2012A&A...541A..51K}, Figure 6).
%it provides a more global view of how the the modes are distributed, which is useful in the case of mixed dipole modes, and it also provides information about the slight non-regularity of the modes, which manifest as curvature in the échelle diagram \cite{2012A&A...541A..51K}.
Still, the three diagrams in concert was useful in the process of visual vetting, particularly for spectra with missing modes, low S/N or full of mixed modes where one diagram alone might be inconclusive. 

\subsubsection{Constructing the training set}
%\subsubsection{Labelling the training set}
\label{sssec:ltts}
Having all three diagnostic plots in place we then start creating the training set.
%For the training set selection, w
To obtain a representative sample we use stars from a wide range of K2 campaigns. We chose the 15,585 stars observed during campaigns 1, 4, 8, 13 and 15, that were deemed to be potentially oscillating giants by the machine learning algorithm from \cite{2018ApJ...859...64H}. 
In order to obtain a series of \dnu\ values that we could label as reliable or unreliable we ran the SYD pipeline on this full sample of 15,585 spectra and removed only those with values %likely to be a mistake of the pipeline 
in the following ranges: 
   \dnu\ $\le 0.3$ \muhz,
   \numax\ $\le 3$ \muhz, and
    \numax\ $\ge 278.8$ \muhz,
%\dnu $\le 0.3$ \muhz, \numax $\le 3$ \muhz, \numax $\ge 278.8$ \muhz, 
but retained all 15,170 remaining stars irrespective of the data actually showing oscillations or not. This was to have a significant fraction of \dnu\ values that we would later label visually as unreliable %(either because there were no oscillations or \dnu\ could not be determined accurately), 
with the aim of having a training set with roughly equal numbers of reliable and unreliable labeled \dnu\ values.% \textcolor{red}{[and thus avoiding the class-imbalance problem that arises because machine learning algorithms normally are trained to minimise the loss function.]}}
%We ran the SYD pipeline on this initial sample of 15,585 spectra and removed those with \dnu <0.3\muhz , \numax <3\muhz \textbf{ or \numax >278.8\muhz}. %or \dnu >\numax .
%In these cases the power is too close to zero or the Nyquist frequency, meaning there was not enough spectral data to create the folded spectrum.
%After this cut, we were left with a total of 15,170 stars. 
We note that the source of the \dnu\ values is not important for our training and subsequent results, as long as the final number of reliable and unreliable \dnu\ values ends up being balanced. In order words, we could have used mock-generated \dnu\ values.% or values from any other seismic pipeline}.

For all the stars in the selected sample, we generated the three diagnostic plots, performed visual checks individually for each star, and from this concluded that
%We generated the three diagnostic plots for all of them, performed visual checks individually for each star, and from this concluded that
a total of 7,240 stars showed oscillation mode structure consistent with a correct \dnu\ measurement, meaning:
\begin{itemize}
    \item[-] the autocorrelation showed peaks at multiples of $\sim$\dnu/2,
    \item[-] the folded spectrum followed the modelled template, and/or
    \item[-] modes $l=0$ and $l=2$ aligned in the échelle diagram.
    \end{itemize}
\noindent Hence, these were labelled as reliable. 
%\textbf{(autocorrelation shows peaks at multiples of $\sim$ \dnu/2, the folded spectrum follows the modelled template, and/or modes $l=0$ and $l=2$ align in the echelle diagram)}, hence are labelled as reliable. 
Meanwhile, 7,143 are labelled as unreliable either because there was absolutely no oscillation signal or signature of \dnu, or because the \dnu\ value was %clearly incorrect.
considered too far off. The latter was typically the case when the \dnu\ value was offset more than $\sim$3\% from the value that would align the échelle ridges. If \dnu\ is off by 3\% or more the ridges in the échelle are significantly slanted (\citealt{2011ApJ...739...13S} Figure 5), the peaks in the autocorrelation function are shifted, and the mode pattern in the folded spectrum gets slightly scrambled. %  and shifted relative to the template.}
The remaining 787 stars could not be confidently classified and were left out from training. 
Examples of stars that we visually classified as having reliable \dnu\ are shown in Figure \ref{fig:sampler}. In Appendix \ref{sec:training2} we show a larger sample of reliable and unreliable \dnu.%\ (Figures \ref{fig:trsetfull} and \ref{fig:unreliable8}). }

\begin{figure}
\centering
\includegraphics[width=0.99\columnwidth]{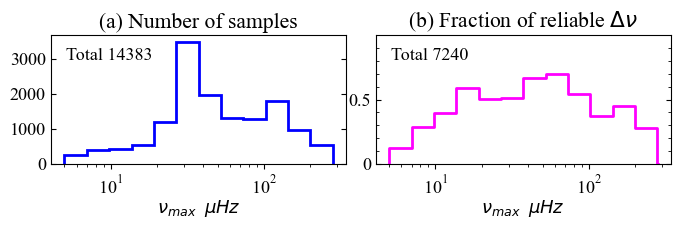}
    \caption{(a) \numax{} distribution of the full training set, (b) Fraction of reliable \dnu{} over the full training set. }
    \label{fig:trset}
\end{figure}

The \numax\ distribution of stars in the training set is shown in Figure \ref{fig:trset}a. 
The fraction of reliable detections as obtained from the described visual method over the totals as a function of \numax\ is shown in Figure \ref{fig:trset}b. For stars with \numax\ below 10\muhz\ the frequency resolution of K2 data makes it difficult to measure and confidently verify \dnu, explaining the lower prevalence of reliable \dnu\ detections.
The lower detection fraction around 30\muhz, where RC stars are typically found, may be caused by their 
%more complicated 
spectra showing a lower height to background ratio (\citealt{2012A&A...537A..30M}, Equation 6) and/or due to their larger number of detectable mixed modes making the spectrum more complex (\citealt{2014A&A...572A..11G}, Figure 7) and hence in both cases harder for the pipelines to find the correct \dnu\ and for us to verify it.
We also see a decline in the fraction of reliable \dnu\ when \numax{} approaches 100\muhz{} and beyond. 
This can be caused by the increased presence of mixed modes throughout the power spectrum and the fact that modes oscillate at lower amplitudes in this \numax{} range leading to lower signal-to-noise ratios.
In the last frequency bin, when \numax\ approaches the Nyquist frequency, we can expect reflections from frequencies greater than Nyquist to interfere with the oscillation pattern, making it harder to identify good \dnu\ measurements. 
The shape of the histogram in Figure \ref{fig:trset}b is similar to that of the six independent pipelines analysed by \citealt{zinn2021k2} (Figure 12). This suggests there is little or no bias unique to our method in the training.

\subsection{Feature Selection}
\label{sec:inputs}

The selection of input features is one of the key concepts in machine learning because the performance of the final model heavily depends on it. 
From our diagnostic plots described in Section \ref{sec:train}
we derive four informative features to provide as input to our machine learning algorithm that mimic what we used for the visual classification of the training set. In this section we describe how we derived each feature.

\subsubsection{Feature AC - based on the Autocorrelation Function}
\label{sssec:inputsAC}
To automate the process of looking for the characteristic peaks in the autocorrelation function
we assign a score to each autocorrelation based on the contrast between the regions where strong correlation is expected and the rest of the function.
\begin{figure}
    \centering
    \includegraphics[width=0.85\columnwidth]{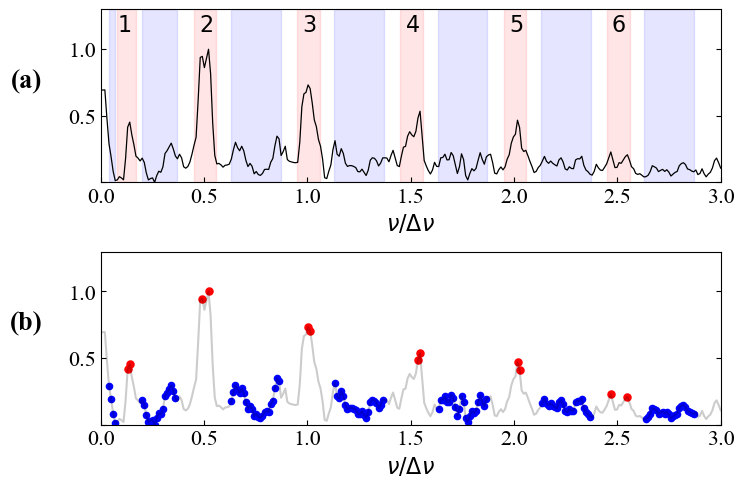}
        \caption{
        Example of calculating a reliability score from the autocorrelation function of EPIC 201207669. a) The detrended autocorrelation of the power spectrum is shown. Annotated red shaded areas mark the six regions of interest, where the autocorrelation is expected to be strongest, and 
        the blue shaded regions where the autocorrelation is expected to be weaker. %The valleys to the left and right of region 3 are annotated, where left(3) = right(2) and right(3) = left(4). 
        b) The final score is obtained as the weighted sum of the local contrast scores, which are calculated for each region of interest as the average of the two highest values in each red region (red dots in the figure) divided by the average of all the points inside the two blue bracketing regions (blue dots).
        }
        \label{fig:acf}
    \end{figure}
The autocorrelation function of star EPIC 201207669 (row II in Figure\ref{fig:sampler}b) is used to exemplify this scoring method. 
We treat the autocorrelation as described in Section \ref{ssec:acf}, but additionally we now fit and remove its background slope using a RANSAC regressor \citep{10.1145/358669.358692}. 
In Figure \ref{fig:acf}a we show the resulting function where red shaded areas indicate our six regions of interest: The first one at $\nu/\Delta\nu=0.12$ indicates the point of expected high autocorrelation from the pairs of $l=0$ and $l=2$ \citep{2010ApJ...713L.176B}.
The rest, at $\nu/\Delta\nu= 0.5, 1.0, 1.5, 2.0$ and $2.5$, indicate the expected peaks from the correlations between $l=0$ and $l=1$ modes. The blue shaded regions to the left and right of each red region are where we expect low correlation, and will be used to derive the contrast.
Figure \ref{fig:acf}b shows, for each of the six red regions, two points with the highest value. 
We derive the contrast score for each red region by dividing the average of the red points by the average of the bracketing blue points.

The final AC score for the star is a weighted sum of these results with weights 
$w = \{0.05, 0.30, 0.50, 0.05, 0.10, 0.00\}$ chosen manually to emphasise the presence of correlation peaks at 0.5 and 1 $\nu$/\dnu, but also to account for correlations at 0.12, 1.5, and 2 $\nu$/\dnu, and adjusted by looking at the performance of different sets of weights for stars with good and bad \dnu.
The weights are modified to $w = \{0., 0., 1., 0., 0., 0.\}$ for low \dnu\ stars (\dnu <1.25\muhz ). For these stars we are only interested in the autocorrelation in region 3, because peaks at \dnu/2 are no longer expected and the pair $l=0,2$ is no longer located at $0.12\nu$/\dnu\ because the oscillations pattern begins to resemble a triplet structure \citep{2014ApJ...788L..10S}.

\subsubsection{Features XC1 and XC2 - based on the Folded Spectrum}
\label{sssec:inputsXC}

We were able to craft a good indicator of the similarity between a star's folded spectrum and its corresponding modelled template from their cross-correlation function. We call this metric XC1, and it measures how the maximum correlation coefficient compares to the rest of the correlation function across all shifts. The procedure to calculate XC1 is illustrated for EPIC 201207669 in Appendix \ref{sec:appendix_metricsXC1}.

We also tried an alternative way of quantifying the similarity between folded spectra and templates: metric XC2 is obtained by calculating the \textit{Manhattan distance} between the model template shifted to the position of maximum correlation 
and a smoothed version of the star's folded spectrum. %, shown in red in Figure \ref{fig:xcexplained}c.
The smoothing is done applying a Butterworth filter of order 4 and cut-off frequency 18 cycles per \dnu, and scaling the result between 0 and 1.
The \textit{Manhattan distance} is the sum of the absolute differences between the observed and modelled folded spectra at each point. 

Even when both XC1 and XC2 aim to extract similar information from the data, we keep them both as inputs for the neural network because a combination of both features is a better \dnu\ reliability indicator than any one of them. (See analysis in Appendix \ref{sec:appendix_metrics}).

\subsubsection{Categorical Feature based on \numax}

There is generally some dependency of how the autocorrelation function looks with \numax\ and therefore adding in \numax\ as a feature is informative. In practice we do this by encoding \numax\ of each star into a categorical feature. %as shown in Table \ref{tab:numax_feats} where the limits are 
The feature is set to define six bins of equal width in logarithmic space between 7.7 and 280 \muhz, where a \numax\ corresponding to the first bin generates the array [1, 0, 0, 0, 0, 0], and so on.

Finally, the \numax\ categorical variable is concatenated with the three features AC, XC1 and XC2 into an array of length 9 for each star. 
This array will be referred to as \textit{Input A}.

\subsubsection{Échelle Diagram as Image Input}
\label{sec:inputB}

Like in the visual labelling process, we found that the neural network performed better 
when adding the échelle diagram as a feature. The validation accuracy during training tests went from roughly 91\% to 94\%.

To create the échelle diagrams for the network we used a 11\dnu{}-wide segment of spectrum around \numax. To process this feature as an image we use a convolutional neural network, which is a special class of deep neural networks particularly useful for image analysis. 
Because échelle diagrams need to be standardised in size before they can be fed to the algorithm, we resize them into 11x150 images using nearest neighbour interpolation. Similarly to the autocorrelation function, the number of columns of the standardised diagram, 150, was chosen conservatively to not introduce smoothing that is too severe for even the narrowest peaks. 
Our choice of using 11 rows, or 11 \dnu\ segments, ensures that all the excess power is always encapsulated in the diagram with at least one row with little or no power on either side of the excess.
The resulting images form what we call \textit{Input B} for our neural network algorithm.

While the échelle diagram implicitly carries all the information described by the other metrics, using it together with the AC, XC1 and XC2 metrics yields the best network performance.

\subsection{Network Architecture}
\label{sec:model_arch}

There is no machine learning architecture that is a priori guaranteed to work better than other for any dataset \citep{10.1162/neco.1996.8.7.1341}, therefore the only way to find the optimal model for a problem would be to evaluate them all. In practice, the way to choose a suitable algorithm for a problem is to make reasonable assumptions about the data and evaluate only a few reasonable models with a limited number of hyperparameters known to work well for similar tasks.
For our classification problem we assume that a deep neural network architecture will be appropriate for \textit{Input A} (Branch A), while we anticipate that a convolutional network (Branch B) will be suitable for our image-like features \textit{Input B}. 
Once reasonable performances are reached, fine tuning of the hyperparameters is not recommended because it can lead to over fitting and hence any improvements in training are unlikely to generalise to new data \citep{geron2019hands}.

\begin{figure}
\begin{center}
\includegraphics[width=0.9\columnwidth]{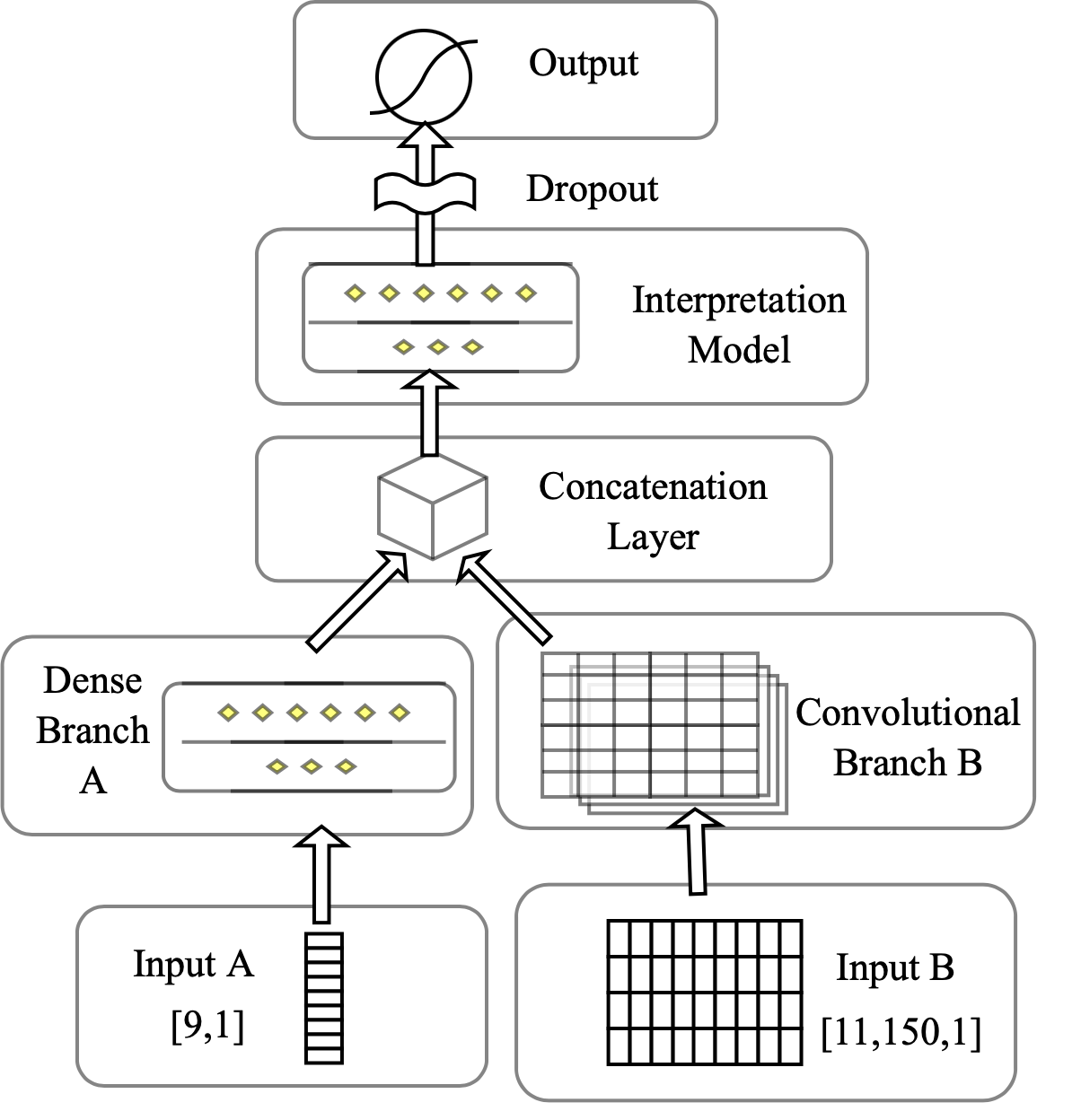}
    \caption{Schematic representation of the Multiple-input algorithm. Input A goes trough a fully-connected neural network (Branch A) and comes out as a 1D array of size [288], and Input B comes out of the convolutional Branch B as a 1D array of size [16], as implied by the parameters detailed in Table \ref{tab:nnparams}. They are concatenated before going through another fully-connected neural network, whose output is of size 1.}
    \label{fig:mim}
\end{center}
\end{figure}

\begin{table*}
    \caption{Neural Network Hyperparameters. We specify the types of layers and the hyperparameters used when defining each layer. "NN" indicates the number of neurons,
    "Activ." indicates the activation function used.}
	\centering
	\label{tab:nnparams}
	\begin{tabular}{ccccc |cccccc|ccccc} 
		\hline
		\multicolumn{4}{|c|}{Dense Branch A} &   & \multicolumn{6}{|c|}{Convolutional Branch B} &   & \multicolumn{4}{|c|}{Interpretation Stage} \\
		%\hline
		Layer & Size & NN & Activ. &  & Layer & Size & Filters & Kernel & Stride & Activ. &  & Layer & NN & Rate &  Activ.\\
		\hline
        Input & [9,1] & - & - & & Input & [11,150,1] & - & - & - & - && Concat. & - & -& -\\
        Dense & &[64] & ReLU&& Conv2D & - & [16]& [3,7] & [1,1] &ReLU && Dense & [64] & - &ReLU\\
        Dense & &[32]& ReLU&& Conv2D & - & [16] & [3,7] & [1,1] &ReLU && Dense & [32] & - & ReLU\\
        Flatten & -& -&  - && Max-Pooling & [1,7] & - & - & - & -  & &  Dropout  & - & 0.75 & - \\
         &&  &&& Conv2D &- &[16] & [3,19] & [1,1] &ReLU& & Dense & [1] & - &Sigmoid\\
        & &  &&& Conv2D & - &[16] & [3,19] & [1,1] &ReLU& & &\\
        & &  &&& {Global} & \multirow{2}{*}{-} & \multirow{2}{*}{-}  & \multirow{2}{*}{-}  & \multirow{2}{*}{-}  & \multirow{2}{*}{-}  & \\
        & &&&&Avg-Pooling& \\
		\hline
	\end{tabular}
\end{table*}
The structure of the neural network algorithm is illustrated in Figure \ref{fig:mim}. 
Outputs from the two branches, A and B, are merged by a concatenation layer and fed to the interpretation stage. Its role is to calculate the relative importance of the results of each branch, which becomes specially important if the two branches return conflicting outputs.
The activation function used until this point in dense and convolutional layers is the Rectified Linear Unit (ReLU) defined as $f(x) = max(0, x)$, where $x$ represents the inputs.
After the interpretation stage we apply a dropout layer \citep{article} with a rate of 0.75. The role of this layer is to randomly and temporarily deactivate 75\% of the neurons from the previous layers during each training step forcing the network to train with a different subset of neurons each time. This is done to prevent overfitting of the training set, which would otherwise happen when the model is optimised for performance on the samples that it has "seen", instead of optimising to generalise on unseen data. Note that the dropout layer is only active during the training phase of the network. 
Outputs from the dropout layer go to the output layer, where a single output neuron with a sigmoid activation function gives the final probability.
The sigmoid function is used generally as an activation function in binary classifiers because it constraints the results to values between 0 and 1, with intermediate values (e.g., 0.5) indicating an uncertain decision.

The hyperparameters of this neural network are summarised in Table \ref{tab:nnparams}. They were a design choice made by manual tuning of multiple combinations of numbers of neurons, filters, dropout rates and kernel sizes. 
Broadly, we followed three guidelines to reach this set of hyperparameters: (a) We required enough complexity (number of neurons) so that the algorithm would converge to a solution. (b) We needed to introduce enough regularisation so that we could train to convergence but before overfitting. 
(c) In Branch B, the kernels in the first couple of convolutional layers were tuned roughly to the size of the features we expect to find in Input B.

Readers interested in learning more about artificial and convolutional neural networks  
are referred to Appendices \ref{sec:ann} and \ref{sec:conv}.

\subsection{Training}
\label{sec:train}

From the 14,383 labelled stars, we use 75\% to train the algorithm, and the remaining 25\% to validate the algorithm's performance.
The fraction of reliable \dnu\ (class "1") to unreliable \dnu\ (class "0") is the same in the training and validation samples.

During training we monitor the accuracy and minimise the binary cross entropy, given by:
\begin{equation} \label{eq1}
-\frac{1}{N} \sum_{n=1}^{N} y_{i} \cdot \log (P(y_{i}) + (1-y_{i}) \cdot \log(1-P(y_{i}))  
\end{equation}

\noindent where $y_{i}$ is the truth label for each star $i$; $P(y_{i})$ is the probability assigned to star $i$ by the network, and $N$ is the number of training samples. 
The training is done using a variant of the stochastic gradient descent algorithm called "Adam" \citep{kingma2017adam}, with a learning rate $\eta$ fixed at 0.001. 
"Adam" aims to minimise the loss function (which is indicative of the error rate) by adjusting the weights of the model iteratively, with the caveat that each step of the gradient descent does not use every example like the stochastic gradient descent does. Instead a mini-batch is randomly selected to train the model in steps. The size of the mini-batch determines how many steps are required to train the model using the entire training sample. One pass through the entire sample constitutes 1 epoch. The model was trained using a mini-batch size of 150.

We trained the algorithm several times. For every training session a new random data split was made and the weights were initiated at random values each time. We terminated the training just before it started to overfit. 
From each training session we saved the weights from the epoch with the best validation performance, and used them to build an ensemble of 39 models where each of them has a validation accuracy of $\sim$94\%.
The final results of our neural network classifier are given by the average across this ensemble. This re-sampling method is known as Monte Carlo cross-validation or repeated training/test splits, and is done to allow for a better use of the limited labelled data while allowing better predictions of how well the model will perform on future samples \citep{Kuhn2013}.

\subsection{Network Performance}
\label{sec:perf}
Our choice to use cross-validation during training implies that there is no one unique validation sample unseen by the algorithm, but this also means that we have simulated a different training distribution in each training session, thus allowing reasonable estimates of model performance on unknown data using the training set.

\begin{figure}
\includegraphics[width=0.64\columnwidth]{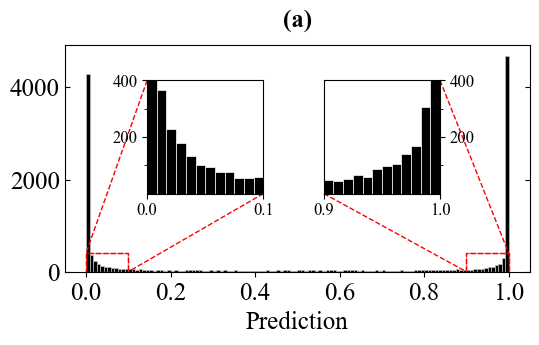}
\includegraphics[width=0.35\columnwidth]{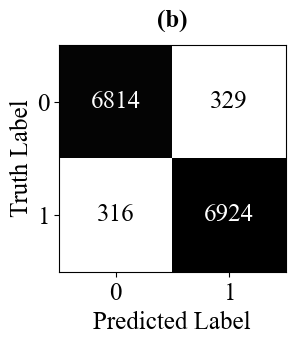}
\caption{(a) Distribution of neural network predictions on the training set. (b) confusion matrix when a threshold is set at $t = 0.5$. Quadrants in black represent the number of correct predictions for each class and the white quadrants the number of mistakes for each class.}
\label{fig:confusion}
\end{figure}

Figure \ref{fig:confusion}a, shows the distribution of the predictions made by the ensemble on the 14,383 labelled stars from the full training sample.
Figure \ref{fig:confusion}b is the confusion matrix when a threshold is set at t = 0.5. Black quadrants represent the correct predictions. The upper-left quadrant shows the number of true negatives, with predictions and truth labels of "0" ("Unreliable \dnu"). The lower-right quadrant shows the number of true positives, the number of stars with predictions and truth labels of "1" (Reliable \dnu). White quadrants show the number of false positives and false negatives: spectra that were predicted to have "Reliable \dnu" (1) when the truth label was "Unreliable \dnu" (0) and vice versa. 

We use the predictions to derive the following three performance metrics: Accuracy, or the number of correct predictions made by the algorithm over all the predictions made, Precision (or purity), or the fraction of correct positives predictions among all the positive predictions made by the algorithm, and Recall (or completeness), or the fraction of correct positive predictions among all the real positives in the data set. Precision is the best metric to optimise when false positives are specifically undesirable, whereas Recall should be optimised when false negatives are specifically undesirable.  

\begin{figure}
\centering
\includegraphics[width=\columnwidth]{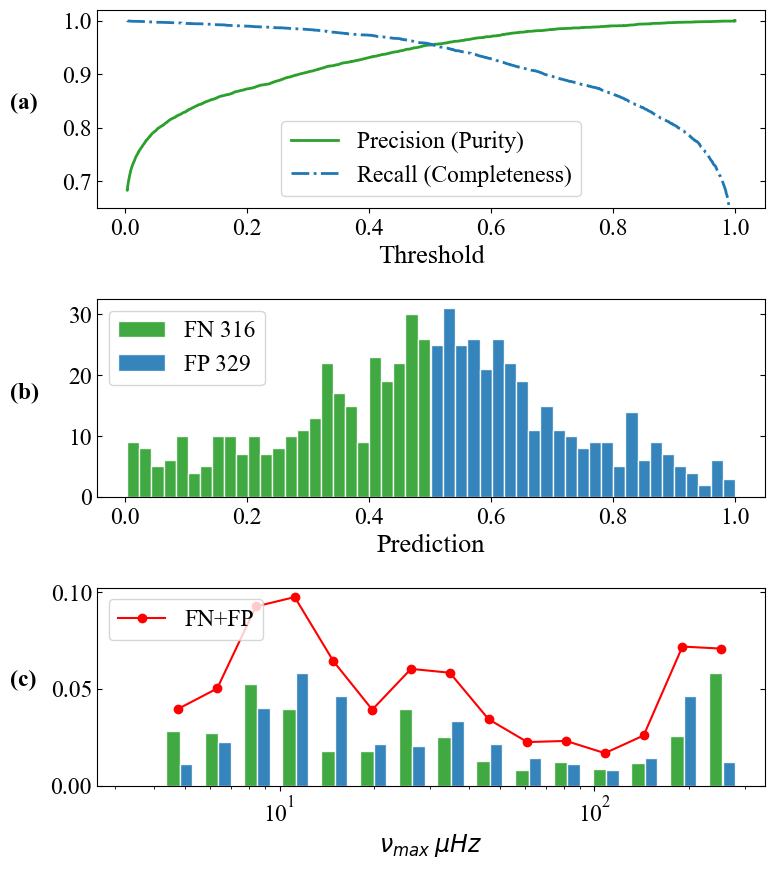}
    \caption{Performance of the neural network classifier on the training set. (a) 
    Precision-Recall (or Purity-Completeness) curve. (b) Prediction distribution of mistakes. (c) \numax{} distribution of the mistakes, normalised to the number of stars from the full training set in each particular bin. In red dots the sum of False Positives and False Negatives for each \numax\ bin. Green and blue bars correspond to FP and FN (which are colour coded as in (b)).}
    \label{fig:perf}
\end{figure}

Figure \ref{fig:perf}a shows the Precision and Recall functions for different probability thresholds \textit{t}. At $\text{probability} = 0.5$, the values of Precision, Recall and Accuracy reach 95.5\%. This is a good decision threshold for us because for our purposes Precision and Recall are equally important.

Figure \ref{fig:perf}b shows the distribution of predicted probabilities for the mistakes made by the network. We find that most of the mistakes are indeed those with intermediate prediction values, as suggested by the fact that in Figure \ref{fig:perf}a, Precision approaches 1 for t>0.8 (meaning false positives approach zero) and Recall approaches 1 for t<0.2 (meaning false negatives approaches zero). 
It follows that to obtain a clean sample with highly reliable \dnu\ and a minimum number of false positives, the threshold can be set higher: for example for $t=0.9$ Precision is 99.66\%. However there would be a trade-off in Recall, which means there will be more false negatives, meaning good \dnu\ values incorrectly vetted out.

Figure \ref{fig:perf}c shows the distribution of incorrect predictions divided by total number of predictions as a function of \numax. Green and blue represent false negative and false positive rates respectively.
The red line shows the combined rate of mistakes. The greatest rates of mistakes occur around stars with \numax{}=10 \muhz{}, which is caused by the low frequency resolution.
There is also a small increase in mistakes around  \numax{}=30 \muhz{} and around \numax{}=200 \muhz{}. We will discuss the challenges of vetting stars in these frequency ranges in section \ref{sec:predict}.

The results just presented tell us that when applying this neural network to a new sample we could expect our outputs to be consistent with its labelling from a human vetter $\sim$95.5\% of the time, granted that the \numax\ distribution of the new sample is similar to the one from this training set.

\section{Results}
\label{sec:results}

In this section we present the results obtained by running our classifier on the K2 GAP sample for which \numax{} and \dnu{} are derived by different pipelines.

In order to evaluate the performance of the classifier on unlabelled data we look for agreement between the \dnu\ and \numax\ for the stars vetted by us and empirically obtained relations from well known oscillating red giant samples observed by \kepler\ \citep{2018ApJS..236...42Y}. We do this first with \dnu{} and \numax{} predictions from the SYD pipeline followed by results from five other pipelines.

\subsection{Results on \dnu\ from SYD pipeline}
\label{sec:predict}

The SYD pipeline is run on all the 
K2 power spectra from campaigns 1-8 and 10-18 corresponding to 47,683 time series (from 45,132 unique targets) that were deemed to potentially show oscillations by the neural network detection algorithm from \citet{2018ApJ...859...64H}. No significance testing or other form of vetting was performed on the resulting \numax\ and \dnu\ results from this SYD run. Hence, by construction we expect a large fraction of \dnu\ values to be incorrect. Less than 20,000 stars are known to actually show oscillations with reliable seismic results for both \numax\ and \dnu\ in the K2 GAP sample \citep{zinn2021k2}. The vetting method that we now implement as part of the SYD pipeline is therefore our neural network classifier, and the resulting vetted SYD values are listed in Table \ref{tab:syd_prob}.

\begin{figure}
\centering
\includegraphics[width=0.98\columnwidth]{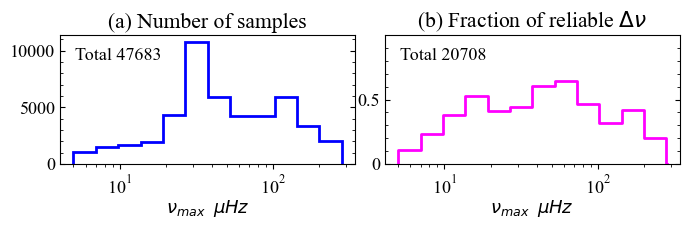}
    \caption{ (a) Distribution of \numax\ for our full K2 sample before vetting, (b) Fraction of good \dnu{} measurements found in every \numax{} bin after vetting the sample from (a).}
    \label{fig:k2gresults}
\end{figure}

Figure \ref{fig:k2gresults}a shows the \numax\ distribution for our entire K2 sample of 47,683 stars, including targets observed during more than one campaign. 
We can see in Figure \ref{fig:k2gresults}b that the fraction of reliable \dnu{} measurements 
found by our vetting has the same general distribution as the corresponding histogram from the training set (Figure \ref{fig:trset}b).

\begin{figure}
\centering
\includegraphics[width=\columnwidth]{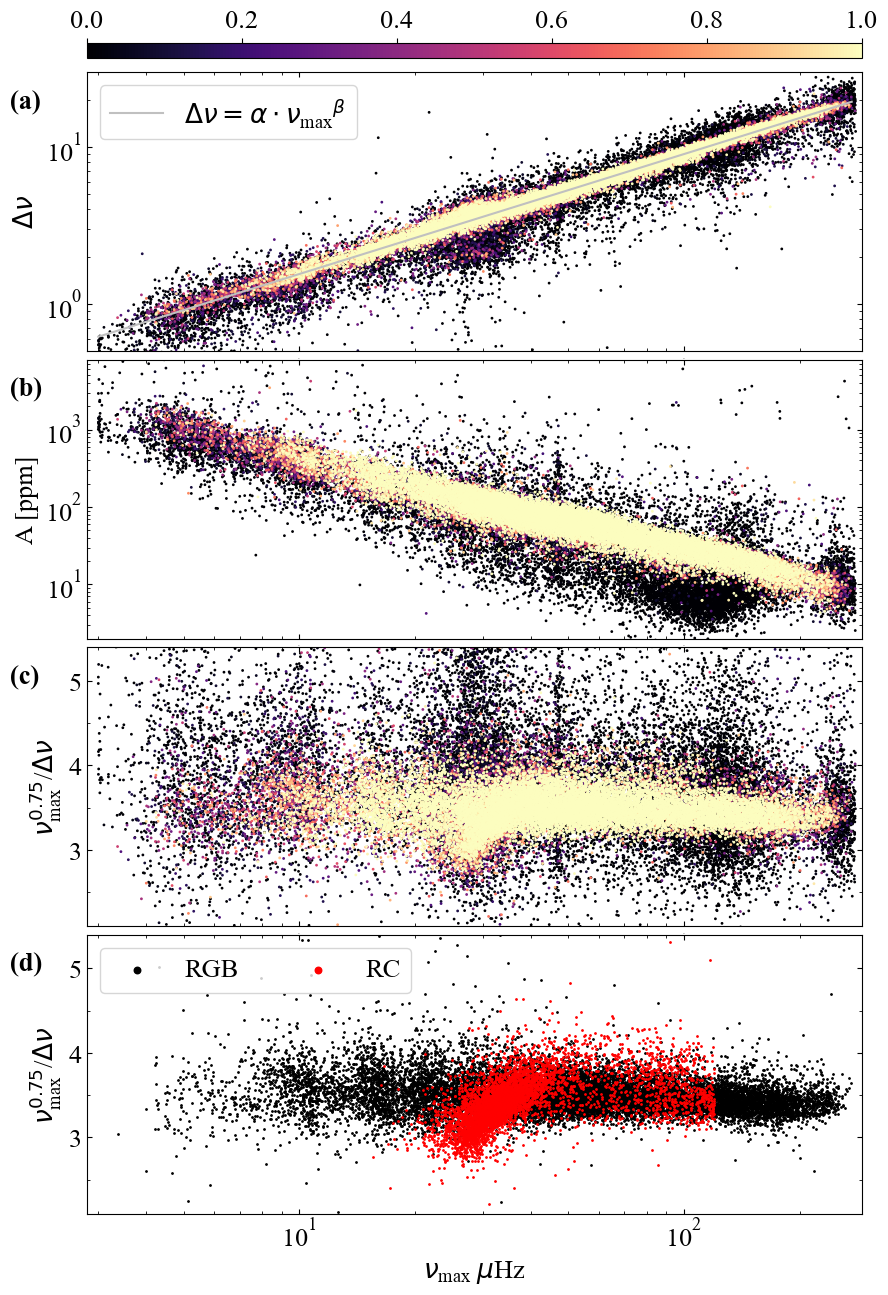}
    \caption{
    SYD pipeline results for the sample of 47,683 time series. Panel (a) shows how the SYD \numax{}-\dnu{} distribution compares to the power law \dnu{}=$\alpha \cdot \nu_{\text{max}}^{\beta}$ where $\alpha=0.26$ and $\beta=0.77$. 
    Panel (b) shows the distribution of oscillation amplitude as given by SYD pipeline with respect to \numax{}. Panel (c) shows the distribution of the asteroseismic proxy for mass with respect to \numax{}. All results on panels a, b and c are colour-coded according to the probability assigned to them by the neural network. Panel (d) shows the same distribution as in panel (c) but only for those stars with probability >0.5 (20,708 observations in total, 19,577 unique targets), and making a distinction in colour based on evolutionary phase. }
    \label{fig:pred}
\end{figure}

To analyse the performance of the automated vetting, we first compare our predictions against the well established power law relation between \numax{} and \dnu{} from \citet{2009MNRAS.400L..80S}, where
 $\Delta\nu = 0.26 \cdot \nu_{\mathrm{max}}^{0.77}$.
 Figure \ref{fig:pred}a shows this \numax{}-\dnu{} relation with a grey line.
The scatter points correspond to the entire sample of 47,683 stars colour-coded by the probability assigned to each star by the neural network. 
The stars in yellow, which indicate high probability of having a good \dnu{}, closely follow the power law.
The stars for which the classifier predicts with certainty that \dnu{} is wrong (darkest points), are those furthest from the power law; these points correspond to measurements that are unphysical.
The points coloured from violet to orange show uncertainty in the predictions (interim values) and are mostly concentrated at either low or high \numax\ or around 30 \muhz, as expected from the discussion in section \ref{sssec:ltts}.

To further verify if our vetting performs as desired, we show in Figure \ref{fig:pred}b the relation between oscillation amplitude and \numax, which is known to follow a power law distribution for solar like oscillators. Particularly, like seen in the \kepler{} sample \citep{2018ApJS..236...42Y}, we see that the stars with a high probability of correct \dnu{} measurements show a sharp upper edge along the power law relation they define.

\begin{figure}
\centering
\includegraphics[width=.85\columnwidth]{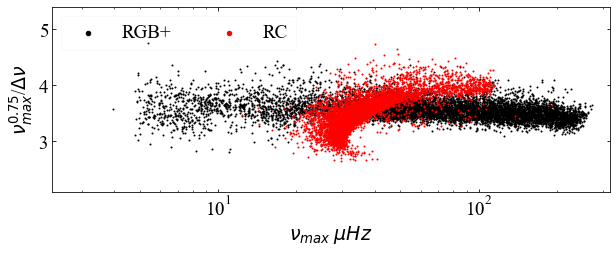}
    \caption{Diagram showing $\nu_{\text{max}} ^{0.75}/\Delta\nu$ as a function of \numax{} for the Kepler sample of 16,000 stars from \citet{2018ApJS..236...42Y} where \numax\ and \dnu\ were derived by the SYD pipeline from 6-month spectra. All \dnu\ values were visually vetted by Yu et al. RC stars are shown in red, in black are RGB stars and stars with no RC/RGB classification available.
    }
    \label{fig:kepler}
\end{figure}

Figure \ref{fig:pred}c shows $\nu_{\text{max}}^{0.75}/\Delta\nu$ as a function of \numax{}, 
where the ordinate is essentially a proxy for mass because: 
\begin{equation}
    \frac{\left(\nu_\mathrm{max}/\nu_{\odot} \right)^{0.75}}{\Delta\nu/\Delta\nu_{\odot}} \simeq 
    \left({\frac{M}{M_{\odot}}}\right)^{0.25} 
    \left({\frac{T_{\text{eff}}}{T_{\text{eff} \odot}}}\right)^{-0.375},
    \label{eq:massnumax}
\end{equation}
\noindent and \teff\ is nearly the same for all giants.
The high probability points in yellow-dominated areas describe the same general shape as the \kepler\ sample shown in Figure \ref{fig:kepler}. 
The excess of scatter points forming a vertical stripe near \numax{}=46 \muhz{} coincide with K2's 6-hour thruster firings.
The \citet{2018ApJ...859...64H} method erroneously flagged these to be oscillations; Here our method can clearly identify and remove these stars whose detected signal is not astrophysical in nature.
Sixty six accepted values of \dnu\ fall beyond the vertical range plotted in panel c (and d). They are mostly false positives with probabilities between 0.5 and 0.8, where \dnu\ given by the pipeline is one half of its real value, and our algorithm was misled by alignment of the wrong modes.

Finally, we show in Figure \ref{fig:pred}d the same diagram as in c but now only for stars with a vetting probability higher than 0.5, and we colour-code stars according to their RGB/RC classification from \citet{2018MNRAS.476.3233H} based on the values of \numax\ and \dnu\ (except for stars with \numax >110\muhz, which we all label as RGB). 
It is reassuring to see the similarity between Figure \ref{fig:pred}d and the corresponding diagram from the \kepler\ sample in Figure \ref{fig:kepler}. 

\subsection{Results on \dnu\ from various pipelines}
\label{sec:pipelines}

In \citealt{zinn2021k2} the authors perform an ensemble-based vetting on \numax\ and \dnu\ for the K2 sample as derived by six automated pipelines\textbf{\footnote{A one-to-one analysis of the different pipelines' performances is presented in \citealt{zinn2021k2}, Section 4.}}: A2Z, BAM, BHM, CAN, COR, and SYD. Their ensemble vetting includes an iterative scaling and averaging process that makes use of results derived by the six pipelines to obtain corrected values of \numax\ and \dnu, but in the latter case excluding A2Z. During the vetting process, \dnu\ values that deviate from the agreement of the rest of the pipelines are clipped out. The final ensemble-vetted sample contains those stars for which \dnu\ from at least two pipelines have `survived' the clipping and contributed to the final corrected value. In this section we will be vetting the original values (unscaled and non-ensemble-vetted) used by \citealt{zinn2021k2} from each pipeline A2Z, BAM, BHM, CAN, and COR. Later we look for differences between our vetted samples and the ensemble-vetted sample in search for our network's mistakes.
Because the SYD sample used in \citealt{zinn2021k2} was  already vetted by our neural network, the SYD data are not treated in this section.

\begin{table*}
    \caption{
    (a) Number of stars before and after our neural network vetting over the same samples used by the ensemble method.
    (b) Number of stars before and after ensemble vetting of each pipeline's results.
    A2Z \dnu\ values are not retained by the ensemble. 
    The 'Before' sample sizes in (a) do not exactly match (b) because the selection process is slightly different for the neural-network and ensemble vetting.  A distinction is made between Non-RC and RC stars. Total numbers are also shown.
    }
	\centering
	\label{tab:pipes}
	\begin{tabular}{cccccccccccccc} 
		\hline
		&  Pipeline && \multicolumn{3}{|c|}{Non-RC} &&  \multicolumn{3}{|c|}{RC} & & \multicolumn{3}{|c|}{Total}\\
		\textbf{(a)}&  Name & & Before & After & Ret\% && Before & After & Ret\% & &Before & After & Ret\% \\
		\hline
        \parbox[c]{0pt}{\multirow{6}{*}{\rotatebox[origin=c]{90}{\textbf{N.N. Vetting}}}} & A2Z && 18,306 & 7,653 & 41.8  && 3,869 & 1,397 &  35.9 && 22,175 & 9,050 &  40.8 \\
        & BAM & &9,031 & 7,767 & 86.0 && 2,490 & 1,900 &  76.3 && 11,521  & 9,667   &  83.9 \\
        & BHM & &16,456 & 12,510 & 76.0 && 5,255 & 3,527 & 67.1 && 21,711  &  16,037  &  73.9 \\
        & CAN & &13,302 & 10,119 & 76.1 && 4,396 & 2,634 &  59.9 &&  17,698 &  12,753  & 72.1  \\
        & COR & &16,823 & 12,351 & 73.4 && 5,192 & 3,726 &  71.8&&  22,015 &  16,077  &  73.0 \\
        & SYD$^{*}$ & &- & 14,620 &  - && - & 4,957 &  - && -  &  19,577  &  -\\
		\hline
		&  Pipeline && \multicolumn{3}{|c|}{Non-RC} && \multicolumn{3}{|c|}{RC}& & \multicolumn{3}{|c|}{Total}\\
		\textbf{(b)} & Name& & Before & After & Ret\% && Before & After & Ret\% & &Before & After & Ret\%\\
		\hline
        \parbox[c]{0pt}{\multirow{6}{*}{\rotatebox[origin=c]{90}{\textbf{Ens. Vetting}}}} & A2Z && 18,331 & - & -  & &3,870 & - & - && 22,201  & -  &  - \\
        & BAM & &9,421 & 7,362 & 78.1 && 2,491 & 2,261 & 90.8 && 11,912  & 9,623  &  80.8 \\
        & BHM & &16,657 & 11,006 & 66.1 & & 5,260 & 4,948 & 94.1 && 21,917  & 15,954  & 72.8  \\
        & CAN & & 13,471 & 9,512 & 70.6 & & 4,397 & 4,085 & 92.9 && 17,868  &  13,597 &  76.1 \\
        & COR & &18,610 & 10,985 & 59.0 & & 5,197 & 4,835 & 93.0 && 23,807  & 15,820  & 66.5  \\
        & ENS & & - & 12,978 & - & &- & 5,843 & - &   &-& 18,821  &  - \\
		\hline	
\multicolumn{14}{l}{$^{*}$Because the SYD pipeline's internal vetting used our neural network vetter, the `Before' and `After' numbers are}\\ 
\multicolumn{14}{l}{{the same. The number of stars for SYD are slightly different to those in \citealt{zinn2021k2} because the latter had}}\\
\multicolumn{14}{l}{{additional cuts applied (see Zinn et al. for details).}}\\
\end{tabular}
\end{table*}

Our vetted results and the probabilities given by the network after running it on  original samples from the five pipelines are listed in Table \ref{tab:pipes_prob}, broken down by campaign. 
The vetting of the samples after removing duplicated stars is summarised in Table \ref{tab:pipes}a, where a distinction is made between RC and non-RC stars as per the results from \citet{2018MNRAS.476.3233H} using ensemble-corrected values of \numax\ and \dnu\ (non-RC stars is our naming for RGB, RGB/AGB, and stars for which an evolutionary classification has not been possible). The %original 
samples used here were already vetted by the five pipelines' own internal methods. The number of those stars are listed as `Before' meaning before our neural network vetting, while `After' refers to after applying our vetting.
For completeness, we have also included in Table \ref{tab:pipes}a our vetted SYD sample from Section \ref{sec:predict} after removing duplicated observations.

\begin{figure*}
\centering
\includegraphics[width=0.75\textwidth]{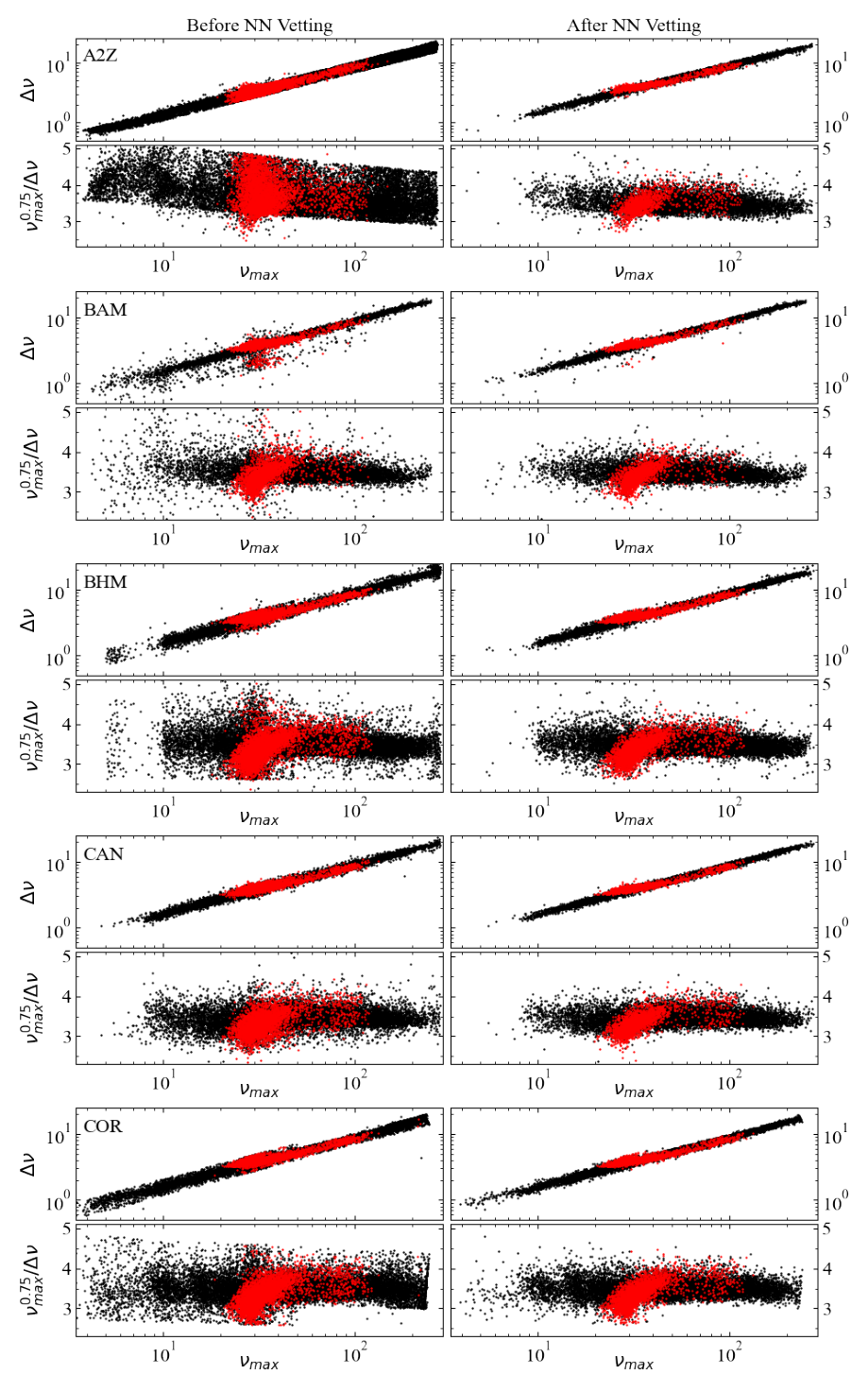}
    \caption{Results previously vetted by each pipeline on the left, and on the right columns we show the sample from the left vetted by our classifier. For each pipeline we show \dnu\ and $\nu_{\text{max}} ^{0.75}/\Delta\nu$ distributions as a function of \numax.
    RC stars appear in red and non-RC stars appear in black. All frequencies given in \muhz. }
    \label{fig:pipes_grid}
\end{figure*}

Figure \ref{fig:pipes_grid} shows the stars from Table \ref{tab:pipes}a before and after our vetting in the form of \dnu\ as a function of \numax\ and the mass proxy as a function of \numax\ (Equation \ref{eq:massnumax}) for each pipeline. RC stars appear in red, and non-RC stars in black.
For all pipelines we see the vetted \numax-\dnu\ plots (right) have been cleaned from almost all outliers compared to the left plots.
In both diagrams our vetting clears out the sharp artificial cuts, which are evident in the `Before' diagrams of all pipelines except the two Bayesian-based algorithms BAM and CAN. 
This cleaning reveals the astrophysical trend seen in the mass proxy vs. \numax\ diagram from Figure \ref{fig:kepler}. 
This includes a more well-defined `hook' of red clump stars, which is an astrophysical feature of low-mass clump stars (\citealt{2010ApJ...723.1607H}, Figure 7; \citealt{2012A&A...537A..30M}, Figure 4).
However, we see a few likely incorrect \dnu\ measurements remaining after our vetting, such as the points remaining from the band under the main \dnu-\numax\ relation for BAM around \numax $\sim$30\muhz. 

\begin{figure}
\centering
\includegraphics[width=\columnwidth]{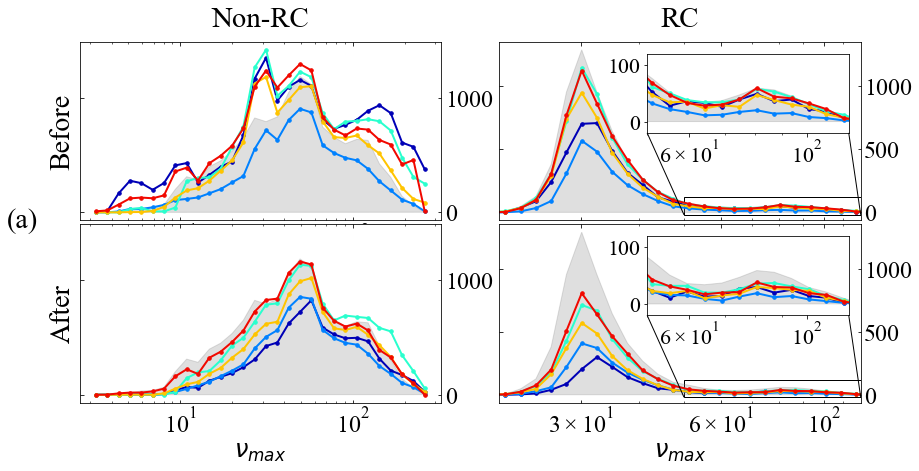}
\includegraphics[width=\columnwidth]{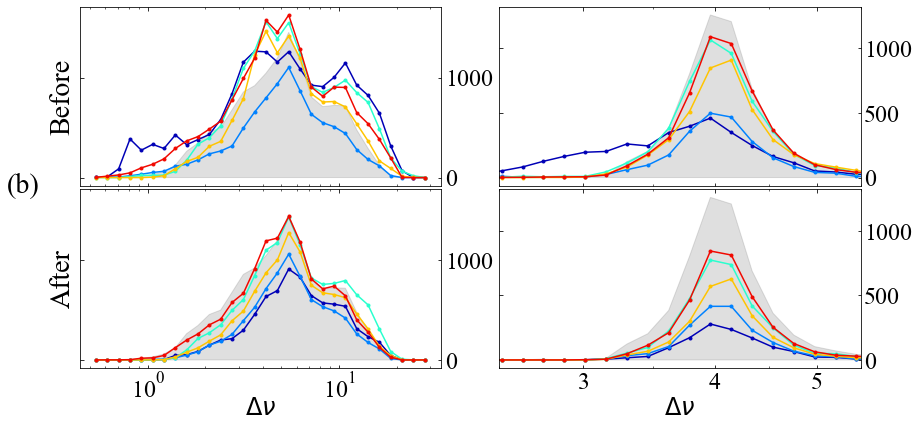}
\includegraphics[width=\columnwidth]{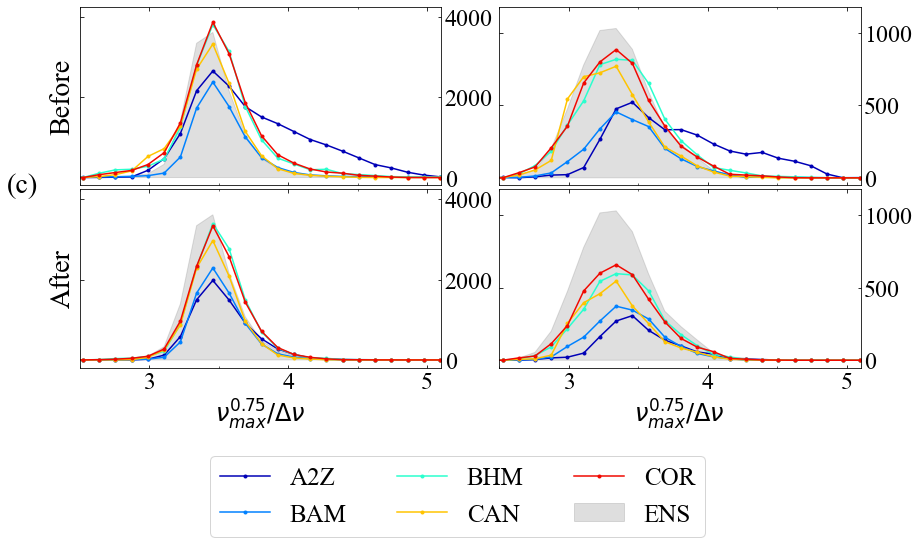}
    \caption{
    Individual pipeline histograms before and after neural network vetting showing (a) \numax\ (b) \dnu\ and (c) $\nu_{\text{max}}^{0.75}/\Delta\nu$ distributions. Left column corresponds to the black points from Figure \ref{fig:pipes_grid} and right column to the red points from the same figure.}
    \label{fig:pipes_RCNRC2}
\end{figure}

In Figure \ref{fig:pipes_RCNRC2} we further investigate the Before/After neural network vetting samples for the different pipelines. The grey filled areas represent the ensemble-vetted RC and non-RC stars, which by construction is the same on the `Before' and `After' rows.
We emphasise that the ensemble is only used for qualitative comparison with the network vetting results because the ensemble method uses corrected values of \numax\ and \dnu\ while our vetting is performed on raw values, and because the two methods are not applied to the exact same samples. 
Yet, our vetting appears successful at removing incorrect \dnu\ values in the case of non-RC stars, which is clear by the fact that lines representing each pipeline `After' vetting are pushed closer towards where the ensemble-vetted results lie compared to `Before' our vetting.

In the following we will examine the differences between our vetted samples and those from the ensemble vetting.
Table \ref{tab:pipes}b shows sample sizes from the different pipelines before and after the ensemble vetting.
The column `After' for each pipeline is the number of \dnu\ values that were used to obtain the final ensemble-corrected \dnu.
Table \ref{tab:pipes} shows that our method retains more non-RC stars than the ensemble method.
However, our neural network is vetting out more RC stars than the ensemble method, and hence possibly removing good \dnu\ values. The lower rate of retained RC stars after our network vetting is also seen in Figure \ref{fig:pipes_RCNRC2}.

\begin{figure}
\begin{center}
    \includegraphics[width=\columnwidth]{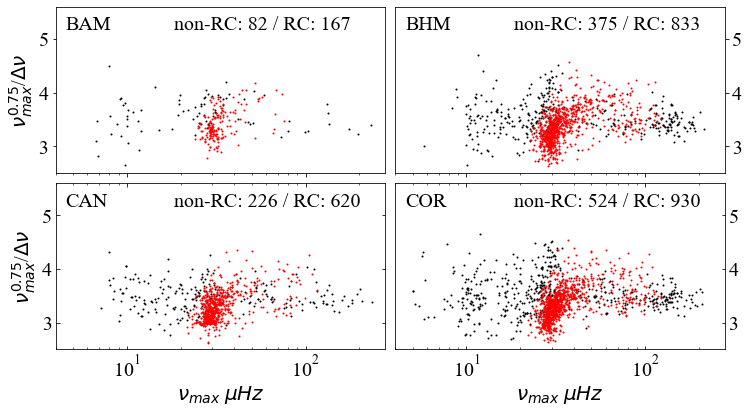}
    \caption{
    Mass proxy diagrams of the suspected false negatives from pipelines BAM, BHM, CAN and COR. From these samples we find many confirmed false negatives, but also unclear/inconclusive cases as well as cases with wrong \numax\ or \dnu\ (real negatives).
    }
    \label{fig:mbypipe2}
\end{center}
\end{figure}

First, we look into those \dnu\ detections that our network removed but were retained by the ensemble vetting. 
We identify for each pipeline all the stars retained by the ensemble method for which our method returns probabilities lower than the threshold of 0.5. 
For many of these, the \numax\ and \dnu\ values differ by a significant amount from their respective ensemble-scaled values. Following our criteria for what we considered a good or bad \dnu\ during the labelling process, we assign as real negatives
those \dnu\ departing 3\% or more from the ensemble-accepted values (See Table \ref{tab:mbypipe}, column 'RN'), while those within 3\% agreement are considered suspected false negatives (Table \ref{tab:mbypipe} `SFN'). The rates of SFN to the total number of \dnu\ analysed from each pipeline are shown as SFN\% for the total number of stars and for RC and non-RC stars separately. Figure \ref{fig:mbypipe2} shows diagrams of $\nu_{\text{max}}^{0.75}/\Delta\nu$ as a function of \numax\ for the suspected false negatives for each pipeline. 
Visual inspection of these stars confirmed many of them as real false negatives. 
We also found unclear cases where visual vetting of the spectra is ambiguous and many cases where the \dnu\ values are offset from a visually-preferred value by $\sim$3\% or more.
So even though these stars are within 3\% from the ensemble-accepted value, we still find quite a few of them not being accurate based on our three diagnostic plots.
A sample of these `true negatives' is presented in Appendix \ref{sec:error}.
The occurrence of these true negatives was most pronounced among the RC stars. This is expected given their less precise ensemble-vetted values, which is evident from the larger spread in RC \dnu\ measurements (compared to RGB) from individual pipelines shown by \citealt{zinn2021k2} in their Figures 10 and 11, bottom left panels. 

\begin{table*}
    \caption{
    \textbf{Stars }
    rejected by our neural network but retained by the ensemble method for each pipeline. Columns 'RN', Real Negatives, indicate those with \dnu\ values departing 3\% or more from the ensemble-scaled values. Columns 'SFN' (suspected false negatives) are those rejected having \dnu\ 
    values within 3\% of the ensemble-scaled value. 
    The rates of SFN to the total number of non-RC, RC, and for all stars analysed by the network are shown as SFN\%.}
	\centering
	\label{tab:mbypipe}
	\begin{tabular}{cc|cccc|cccc|ccc} 
		\hline
		Pipeline && \multicolumn{3}{|c|}{Non-RC} & & \multicolumn{3}{|c|}{RC}&& \multicolumn{3}{|c|}{Total}\\
		Name && RN  & SFN & SFN\% & & RN & SFN &  SFN\% & & RN & SFN &  SFN\%\\
		\hline
        BAM && 238 &  82 & 0.9\% && 255 & 167 & 6.7\% && 493 & 249 & 2.1\% \\
        BHM && 432 & 375 & 2.3\% && 703 & 833 & 15.8\% && 1,135 & 1,208 & 5.6\% \\
        CAN && 375 & 226 & 1.7\% && 894 & 620 & 14.1\% && 1,269 & 846 & 4.8\% \\
        COR && 386 & 524 & 3.1\% && 361 & 930 & 17.9\% && 747 & 1,454 & 6.6\% \\
		\hline
	\end{tabular}
\end{table*}

\begin{figure}
\centering
\includegraphics[width=.85\columnwidth]{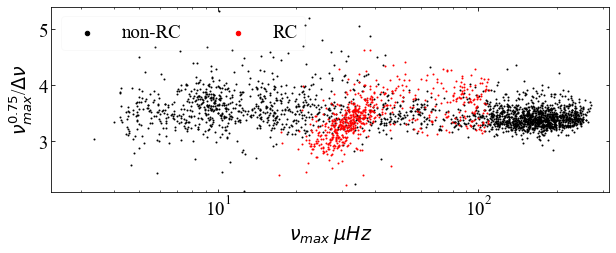}
    \caption{Mass proxy diagram showing SYD parameters for stars vetted by our method and clipped out of the ensemble vetted sample: 2,213 non-RC and 649 RC. These were suspected false positives, but upon visual analysis of a random sample they were found to be true positives (good \dnu) in more than 97\% of the cases.} 
    \label{fig:missing}
\end{figure}

Moving on from stars that our vetting removed (but the ensemble-vetting did not) we now want to examine stars that our vetting preserves but that the ensemble removes. We can see such stars in the `After' plots of non-RC stars in Figure \ref{fig:pipes_RCNRC2}, where our vetting shows a significantly larger number of accepted \dnu\ values from the BHM sample at \numax $\gtrsim$100\muhz\ and \dnu$\gtrsim$10\muhz, and from BHM and COR for ($\nu^{0.75}_{\text{max}}$/\dnu)$\gtrsim$3.5, hence preserving stars that are taken out by the ensemble method. To further examine cases like these, we plot in Figure \ref{fig:missing} the mass proxy diagram for all the stars found in our vetted SYD sample that were clipped out of the ensemble-vetted sample (2,862 stars in total). Visual inspection of this sample indicated that more than 97\% were genuine oscillators, translating into less than 80 false positives (or less than 0.1\% of the total number of stars analysed). This suggests that the ensemble vetting may be removing a significant fraction of genuine oscillators, specially at \numax$\gtrsim$100\muhz. 
Examples of these `extra' stars found to have reliable \dnu\ values are presented in the Appendix, Figure \ref{fig:sclump4}.

Overall, the analysis of our vetting against the ensemble-vetted sample does not contradict the classifier's performance metrics from Section \ref{sec:perf}. The four-pipeline-average of the total rates of suspected false negatives from Table \ref{tab:mbypipe} is 4.8\%, which is slightly more than expected from our initial validation. However, it was found that not all of suspected false negatives correspond to mistakes of the network, as this group also contains \dnu\ with errors of $\sim$3\% or more, and stars with unclear oscillation status upon visual verification.
The real false positives, on the other hand,  were found to be a fraction of a percent of the total number of stars analysed.

\section{Limitations and Biases}
\label{sec:bias}

Our classifier currently shows a higher incidence of false negatives in RC stars. This is most likely explained by our XC1 and XC2 metrics (Section \ref{sssec:inputsXC}), which are based on RGB models and thus may be too harsh in rejecting \dnu\ in RC stars. A future version of this classifier that includes templates for RC stars could help improve our vetting by reducing the rate of false negatives, thus increasing our method's accuracy.

Another limitation of our classifier stems from the difficulties in visually identifying \dnu\ reliably for certain \numax\ ranges. As discussed in Section \ref{sssec:ltts}, this limitation could be explained by the low data resolution in the case of \numax\ lower than $\sim$10 \muhz, and by the lower S/N and the Nyquist frequency mirroring effects in the case of \numax\ higher than $\sim$200 \muhz.
We note from Figure \ref{fig:pipes_grid} (`Before') that different methods show different efficiencies in determining \dnu\ at these \numax\ ranges.
While most stars in these ranges are removed by our vetting, it is evident from Figure \ref{fig:mbypipe2} that almost all of these network-removed stars were also removed by the ensemble method.
This means there was no consensus on the \dnu\ value for the stars vetted out by the ensemble method, indicating that these limitations at low and high \numax\ are intrinsic and not a unique bias to our vetting.
However, because our neural network was trained on data that was manually labelled, a `human' bias is inevitably present.

Despite these limitations, our classifier avoids the drawbacks of vetting methods that employ sharp parameter cuts (undesirable for population analyses) and it provides 
an efficient way to remove outliers in \dnu. In Appendix \ref{sec:vetting} we present an attempt to vet the samples using cuts to the uncertainties in \dnu\ delivered by each pipeline, which demonstrates that the uncertainties for individual stars is not a good measure to identify outliers.

\section{Conclusions}
We have presented a new automated method that efficiently vets asteroseismic \dnu\ measurements applying criteria based on the visual inspection of the spectra as defined in Section \ref{sec:training}. Our 
automated vetting is fully independent of the method used to derive \numax{} and \dnu\ and does not rely on any prior knowledge of empirical relations
as used by many pipelines to constrain \dnu\ detections based on \numax\ \citep{2011A&A...525A.131H}. Furthermore, raw outputs of our classifier can be read as probabilities, demonstrated by the fact that hardly any of the mistakes of the network correspond to high certainty results, i.e. very close to either 0 or 1 (Figure \ref{fig:perf}b.)

From labelled training set performance we see that our neural network is expected to agree with human-vetted samples about 95\% of the time, assuming the \dnu\ distributions of those samples are similar to the one used in our training. 

We tested our results against trusted values from the \kepler\ sample and against K2 ensemble-vetted results.
When applied to pre-vetted samples from five different pipelines we saw that the neural network removed almost all outliers from the diagrams mass-proxy vs. \numax\ and \dnu\ vs. \numax, revealing astrophysical trends expected from the oscillation parameters of solar-like oscillations.
In raw values from four pipelines we found a large number of suspected false negatives: \dnu\ values vetted out by the network but accepted by the ensemble method. Manual checks confirmed false negatives, but also revealed many \dnu\ values with errors larger than $\sim$ 3\% whose rejection is by design of the training sample (Section \ref{sssec:ltts}). 
We also saw a 
higher incidence of false negatives from RC stars when compared to non-RC stars, which had not been previously detected. A future version of the classifier with improvements to RC performance
is planned to be made available to the community, such that it could be added as a last step to any algorithm that measures \dnu. 
When used on the un-vetted sample from SYD we found that the neural network correctly accepted a significant number of stars that the ensemble vetting of the K2 GAP sample is discarding, especially stars with \numax$\gtrsim$100\muhz.

Overall our method appears very promising for fully automated and fast vetting of \dnu\ measurements on large samples of stars as expected from missions like TESS (as applied by \citealt{2021arXiv210705831S}) and PLATO.

%\section*{Acknowledgements}

%...

%%%%%%%%%%%%%%%%%%%%%%%%%%%%%%%%%%%%%%%%%%%%%%%%%%
\section*{Data Availability}
The neural network vetted results are presented in Appendix \ref{sec:tables} and are available as supplementary material.%at MNRAS online.



\bibliographystyle{mnras}
\bibliography{classifier} % if your bibtex file is called example.bib

%\begin{thebibliography}{99}
%\end{thebibliography}

%%%%%%%%%%%%%%%%%%%%%%%%%%%%%%%%%%%%%%%%%%%%%%%%%%

%%%%%%%%%%%%%%%%% APPENDICES %%%%%%%%%%%%%%%%%%%%%

\appendix

\section{Neural Networks}

\subsection{Artificial Neural Networks}
\label{sec:ann}

\begin{figure}
\centering
\includegraphics[width=0.6\columnwidth]{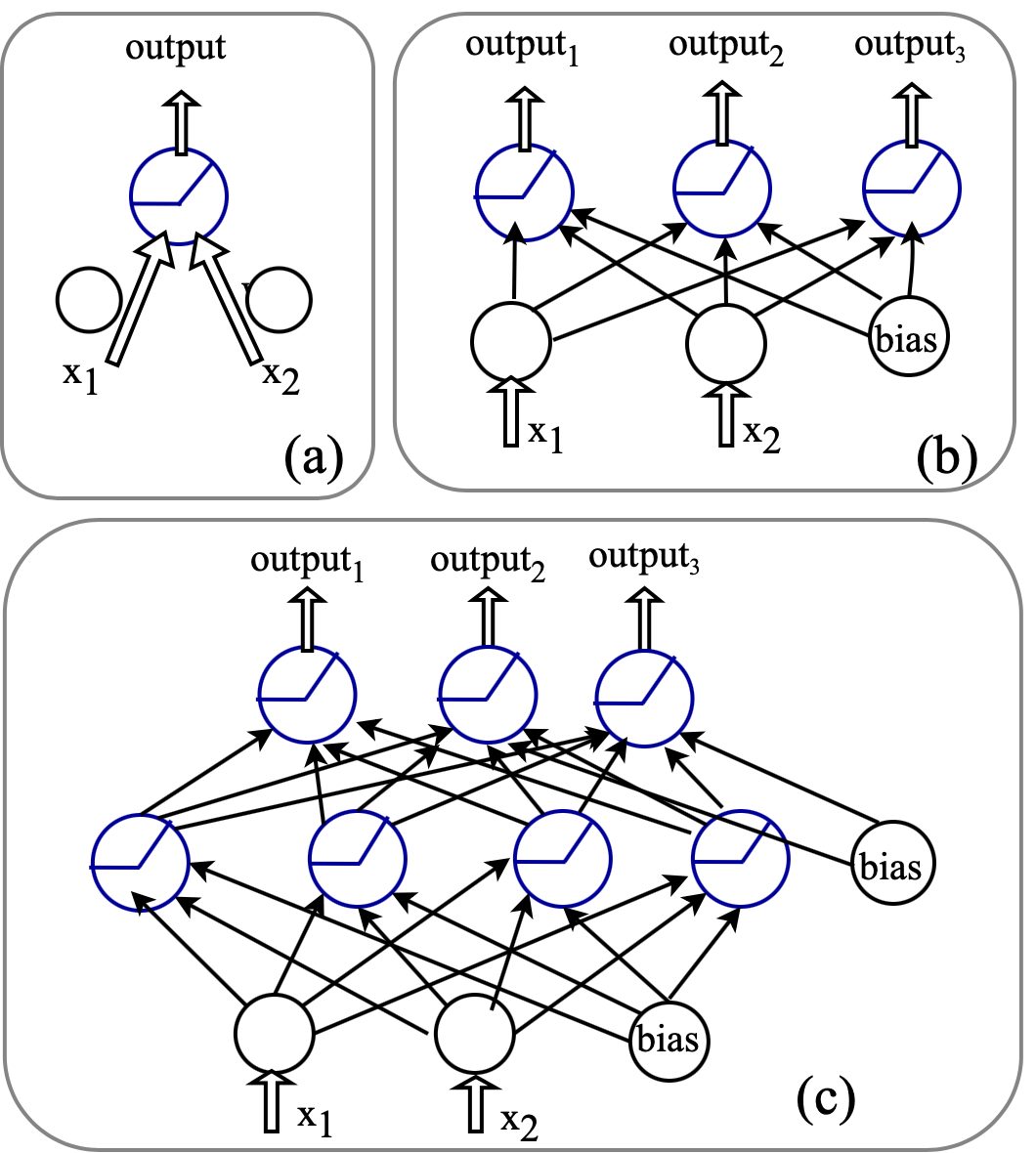}
    \caption{Basic structure of artificial neurons. (a) represents a threshold logic unit (TLU). (b) the perceptron architecture. (c) the multi-layer perceptron architecture MLP with one hidden central layer. Each arrow connecting neurons from different layers has a trainable weight associated (not represented here). Bias neurons represents a constant term that provides flexibility to the results of the network.}
    \label{fig:ANN}
\end{figure}

\begin{figure}
\centering
\includegraphics[width=0.9\columnwidth]{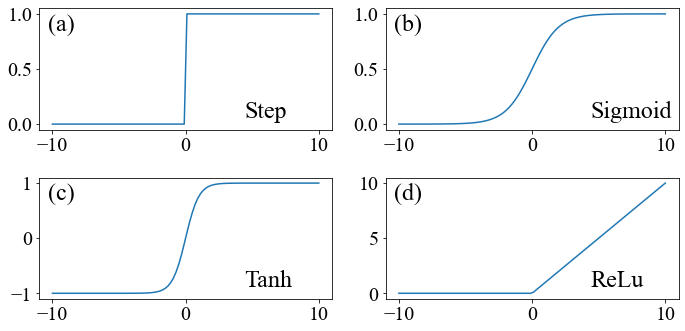}
    \caption{Examples of activation functions (a) A step function simply outputs '0' if its input is negative and '1' otherwise. (b) the sigmoid function: $f(x)=\frac{1}{1+e^{-x}}$, (c) Tanh: $f(x)=\frac{2}{1+e^{-2x}}-1$ and (d) ReLU:  $f(x) = 0$ if $x<0$ and $f(x) = x$ if $x>=0$. In our model (Figure \ref{fig:mim}) we used a sigmoid function for the output layer, and    rectified linear unit (ReLU) functions in the dense layers. }
    \label{fig:activation}
\end{figure}

An artificial neuron called a threshold logic unit is shown in Figure \ref{fig:ANN}a. The threshold logic unit is a simple mathematical unit that produces an output signal by applying an activation function $\phi$ over a weighted sum of its inputs $\vec{X}$ (in our examples $\vec{X}=[ x_{1}, x_{2}]$ ). Its output can be expressed as $\phi(\vec{X}^T \vec{W})$ where $\vec{W}$ represents the weights associated to each connection, which are shown graphically as arrows in Figure \ref{fig:ANN}.
Activation functions, also known as transfer functions, define how the weighted sum of the input is transformed into an output by mapping that sum to a predefined set of values. Examples of activation functions are shown in Figure \ref{fig:activation}.
In practice, a network of neurons as shown in Figures \ref{fig:ANN}b and \ref{fig:ANN}c, is used to increase the complexity of functions that can be modelled or estimated with neurons.

The Perceptron shown in Figure \ref{fig:ANN}b is a network comprising a single layer of neurons.
Graphically, we can describe the neuron layer in a Perceptron as a \textit{row} of neurons, which has connections to the neurons in layers above and/or below it but not within the row itself.
Perceptrons are trained by updating their weights using the equation $w=  w^{\prime}+ \eta (y - \hat{y}) x$.
During each update the connecting weights of every neuron in the network
are re-calculated by adding to the current weight $w^{\prime}$ the difference between the expected output $y$ and the output obtained in the previous step $\hat{y}$, multiplied by a learning rate $\eta$ and its input $x$. The weights are updated as many times as there are training instances available. This is the basis of the gradient descent algorithm, widely used in machine learning.

In practice, additional layers are stacked to form a Multi-Layer Perceptron as shown in Figure \ref{fig:ANN}c. A Multi-Layer Perceptron is capable of more complex classification tasks due to the greater degree of non-linearity from more layers between input and output. Multi-layer perceptrons are examples of Deep Neural Networks.

\subsection{Convolutional Neural Networks}
\label{sec:conv}

Convolutional neural networks - ConvNets - are the natural choice for machine learning tasks involving images because they are able to capture the spatial dependencies in them through the application of convolving filters.

\begin{figure}
\begin{center}
\includegraphics[width=.9\columnwidth]{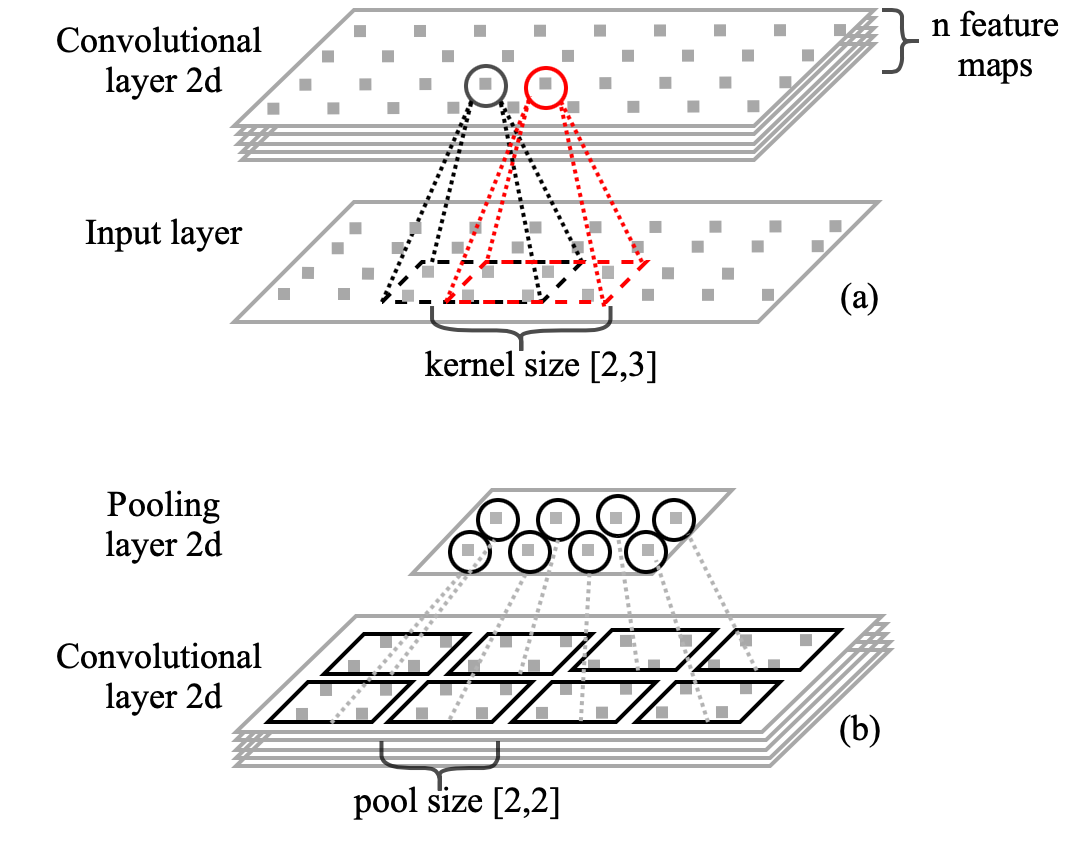} 
    \caption{Two types of layers in a two dimensional convolutional neural network. (a) A convolutional layer of stride=[1,1]: stride is the length (in units of neurons) to skip between adjacent kernels. Here the convolutional layer has the same dimensions as its input layer, and is made of 'n' feature maps. This diagram shows these neurons do not form fully connected layers as only specific neurons of the input layer within a region that is defined by the kernel size can connect to a particular neuron in the next layer. These regions (e.g., black and red), however, can overlap.
    (b) A pooling layer performs an operation on groups of neurons, with the size of the group determined by a pooling size. In this example, the pool size is 2x2 and a maximum operator is applied. This means that within a group of 2x2 neurons, only the neuron with the maximum value is passed to the next layer (max pooling).}

    \label{fig:layers}
\end{center}
\end{figure}

Inspired by neurons in the visual cortex \citep{Fukushima2004NeocognitronAS}, individual neurons in a ConvNet respond to information from only a restricted region of the input image known as their Receptive Field. The overlapping receptive fields corresponding to individual neurons cover the entire visual area.
There are several elements involved when designing ConvNets: 
First, a \textit{convolutional layer} made of n \textit{feature maps}, convolves over the input layer one small sector at a time. The size of the receptive field is known as the \textit{kernel size} as shown in Figure \ref{fig:layers}a. 

All the feature maps in the same convolutional layer share the same kernel size, and all the neurons in the same feature map share the same set of weights. In ConvNets the sets of weights are called \textit{filters} and can be represented as small images the size of the kernel that extract patterns from the inputs. 
Because weights in ConvNets form filters, training such network involves learning image filters that are best suited to perform a particular task.

\textit{Pooling layers} are used as shown in Figure \ref{fig:layers}b. A pooling layer outputs a predefined function of the neurons in its receptive field called the \textit{pool size}. Often the predefined function is the maximum value ('max-pooling') or the average value ('average-pooling'). 

All neural networks need to be designed according to the problem they are trying to solve.
For ConvNets this involves using kernel sizes appropriate to the size of the features we wish to detect in an image. 
It is also important to experiment with different numbers of layers and feature map elements in order to find an optimal structure that delivers good performance while avoiding unnecessary computations.

\section{Examples from the Training Sample}
\label{sec:training2}

\begin{figure*}
\includegraphics[width=0.97\textwidth,height=14cm]{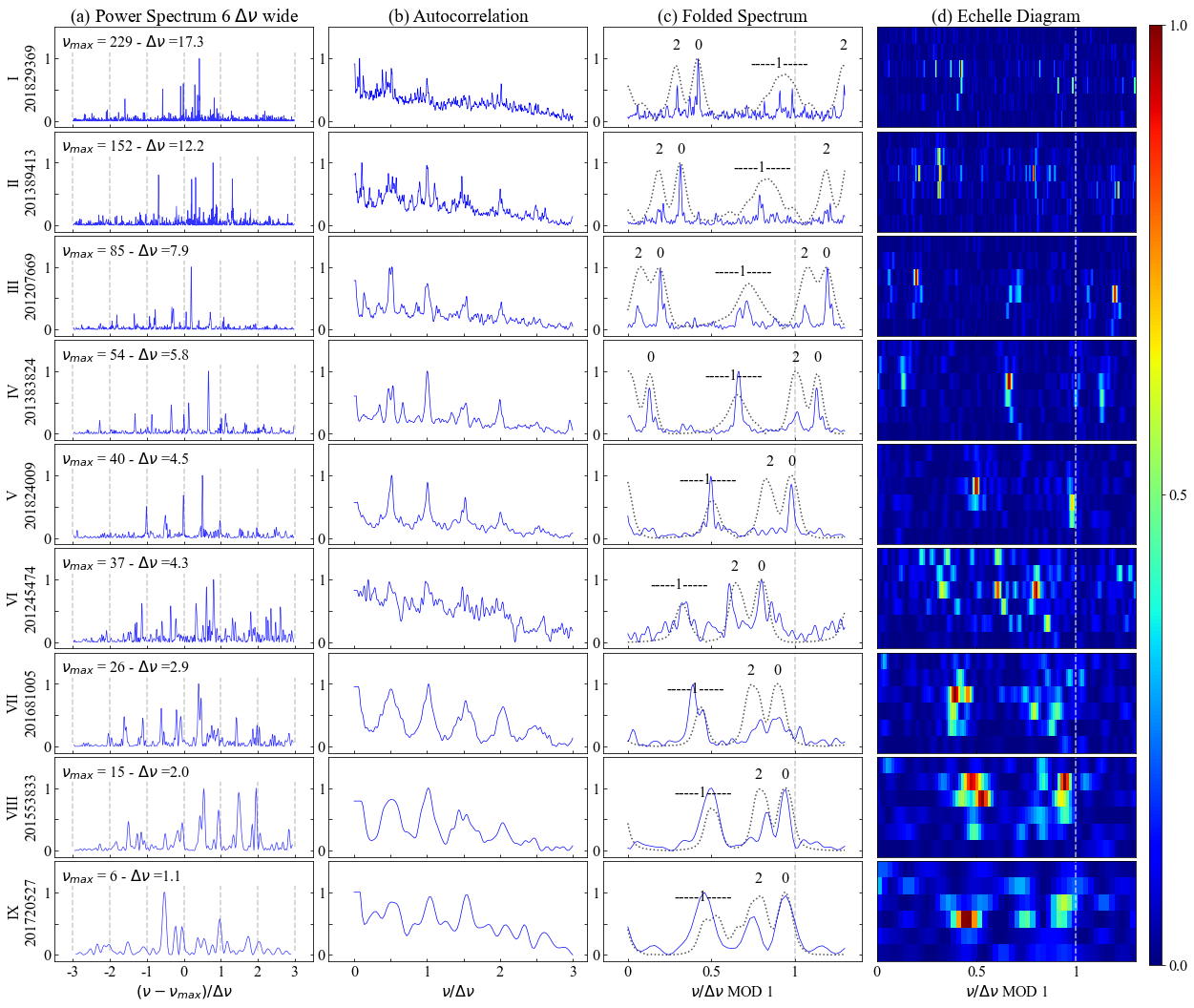}
    \caption{
    Diagnostic plots for a representative set of stars with reliable \dnu\ spanning the full range of \dnu\ values in the K2 sample. Values of \numax{} and \dnu{} are given in \muhz{} and all data have been scaled between 0 and 1.}
    \label{fig:trsetfull}
\end{figure*}

\begin{figure*}
\includegraphics[width=0.97\textwidth,height=14cm]{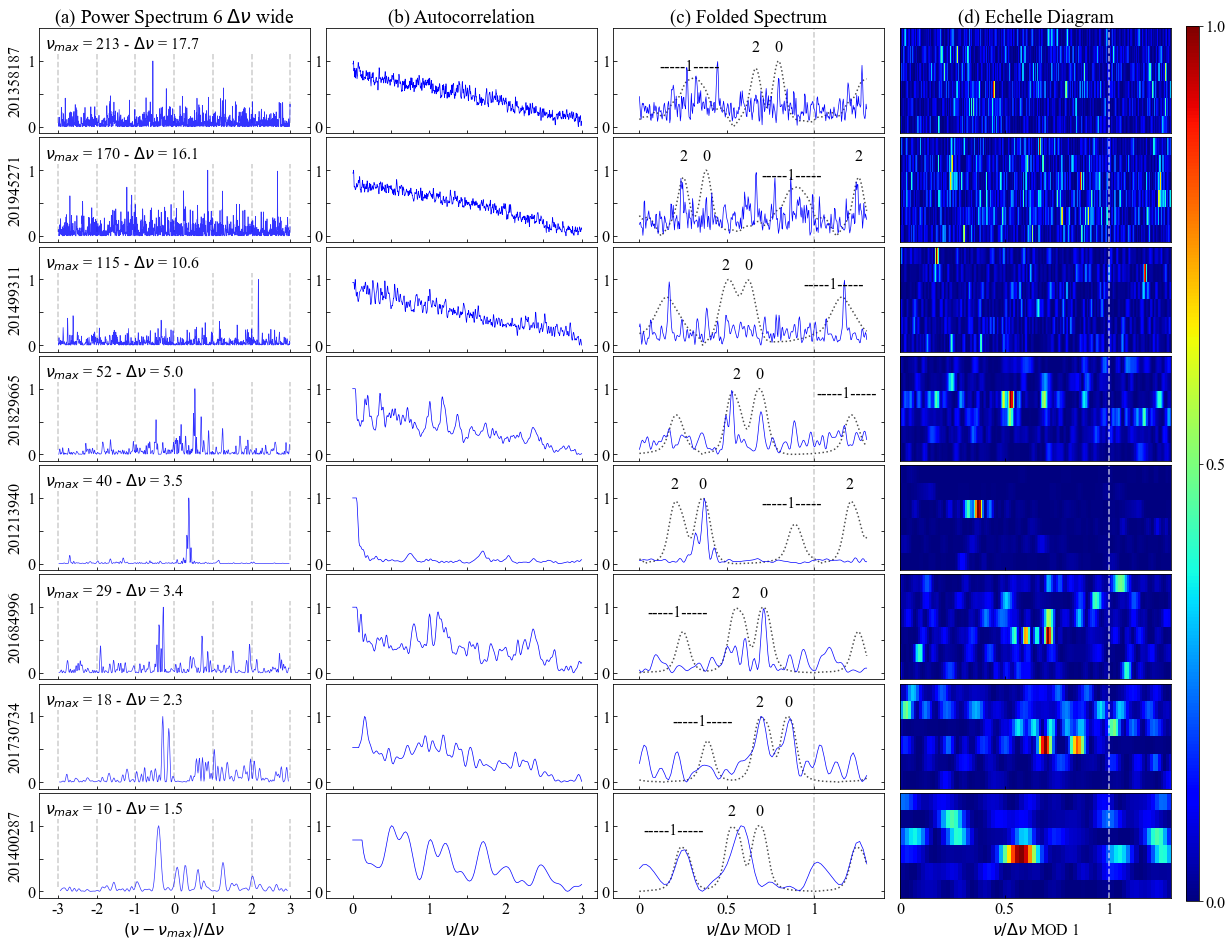}
    \caption{
    Diagnostic plots for a representative set of stars with unreliable \dnu\ spanning the full range of \dnu\ values in the K2 sample. Values of \numax{} and \dnu{} are given in \muhz{} and all data have been scaled between 0 and 1.}
    \label{fig:unreliable8}
\end{figure*}

Figure \ref{fig:trsetfull} shows examples of stars with reliable \dnu{} from our training set. They represent red giants in different evolutionary phases from the bottom to the tip of the red giant branch, except for EPIC 201245474 in row VI, which appears to be a RC star because there are many mixed modes appearing all over the spectrum with heights similar to the acoustic modes.
These examples are representative for our K2 sample. 
For high \numax\ frequencies (rows I and II) %we see mixed modes specially around l=1. 
the presence of mixed modes is evident in diagrams b, c and d. For intermediate \numax{} (examples from rows III and IV) we do not see many mixed modes.  %, which makes it easier to identify the p modes \textbf{visually}.
A feature of our K2 sample is that occasionally some modes appear to be missing due to the stochastic nature of the oscillations and the relatively short duration of the light curves. The example in row V demonstrates this: $l=2$ seems to be missing, but there are small peaks around the  0.5\dnu\ and 1\dnu\ main peaks in the autocorrelation. This indicates the quadrupoles are there, only with lower power than normal.
Continuing down the list, we see that peaks appear wider; this is because the frequency resolution is the same while the frequency separation becomes smaller. 
The last example in row IX is one of the clearest examples of oscillations found in this low frequency range. It shows that the autocorrelation function becomes broader and that there are only a couple of orders with power.  
In contrast, Figure \ref{fig:unreliable8} shows examples of spectra with unreliable \dnu{} from our training set, which are the ones that our method aims to remove.

\section{Metrics}

\subsection{Calculating metric XC1}
\label{sec:appendix_metricsXC1}
\begin{figure}
    \centering
    \includegraphics[width=.9\columnwidth]{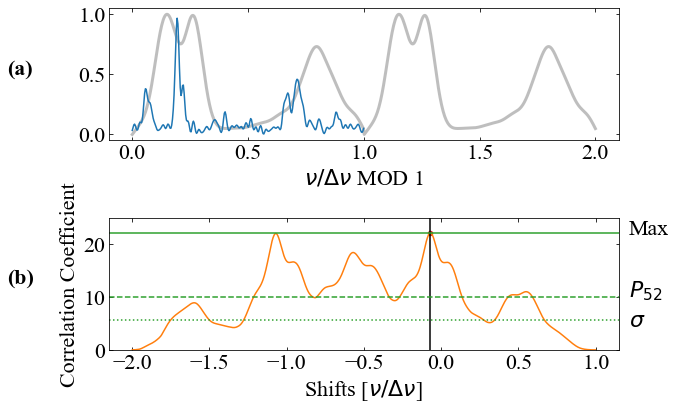}
    \caption{ Example of calculating metric XC1 for EPIC 201207669 based on the correlation between its folded spectrum and the folded template (model C).
    (a) the star's folded spectrum in blue and two copies of the folded template in grey. (b) the full correlation of the functions where Shift=0 corresponds to their relative position as shown in (a). The position of maximum correlation is marked with a black vertical line. The value of the maximum correlation coefficient, the 52nd percentile of the entire function and its standard deviation are marked with solid, dashed, and dotted horizontal green lines, respectively. XC1 is then the difference between the maximum correlation coefficient and the 52nd percentile divided by the standard deviation.
    }
    \label{fig:xcexplained}
\end{figure}

We describe the procedure used to calculate metric XC1 that quantifies the similarity between each star's folded spectrum and the folded template obtained from 1 \msol\ models, as described in Section \ref{sec:FS}. Figures \ref{fig:xcexplained}a and \ref{fig:xcexplained}b illustrate this for EPIC 201207669: first we chose the folded model template for this star's \dnu\ of 7.9\muhz, which is model C according to Table \ref{tab:table_models}. Figure \ref{fig:xcexplained}a shows the star's folded spectrum in blue and two copies of the template in grey. Figure \ref{fig:xcexplained}b shows the full correlation between the functions from panel (a). We create feature \textit{XC1} by subtracting the 52nd-percentile of the correlation (green dashed line in Figure \ref{fig:xcexplained}b) from the maximum correlation (solid green line), and divide this result by the standard deviation (green dotted line). We tried several different percentiles to calculate this indicator, and found that the 52nd-percentile resulted in the best separation of good and bad \dnu\ values in the training set. However, a similar performance could be obtained if choosing values between the 45th and the 60th percentile.

\subsection{Performance of the three metrics}
\label{sec:appendix_metrics}
\begin{figure}
    \centering
    \includegraphics[width=\columnwidth]{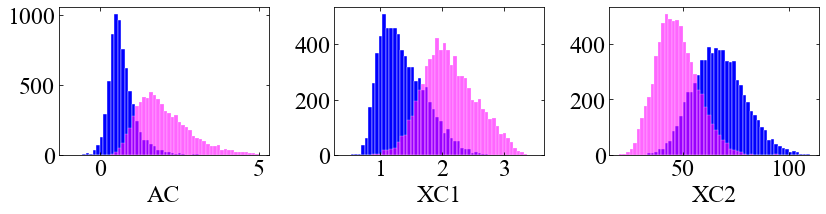}
    \caption{
    Distribution of the numerical metrics AC, XC1, and XC2 shown in magenta for stars in the training set with good \dnu\ and in blue for the stars with bad \dnu .}
    \label{fig:metrics}
\end{figure}

\begin{figure}
\begin{center}
    \includegraphics[width=\columnwidth]{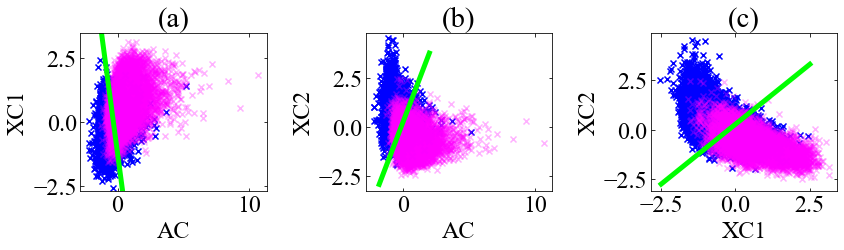}
    \caption{Scatter plots describing the distribution of the stars from the training set over pairs of the metrics described in Section \ref{sec:inputs} (after standardisation of each metric). Blue crosses correspond to unreliable \dnu\ and magenta crosses are reliable \dnu\ from the training set. Green lines are given by the fitting of a linear classifier to the data in the plot.}
    \label{fig:metrics_ap}
\end{center}
\end{figure}

By using histograms we assess the individual ability of the metrics from Subsections \ref{sssec:inputsAC} and \ref{sssec:inputsXC} to separate reliable \dnu\ from unreliable \dnu\ in the training set of 14,383 stars. This is shown in Figure \ref{fig:metrics} where good (reliable) \dnu{} appear in magenta and bad (unreliable) \dnu\ in blue.
The metric with best separability is AC because the intersection of good and bad \dnu\ distributions is only 1,796 stars. This translates into 87.5\% accuracy if we set a threshold at the point where both histograms (blue and magenta) have close to the same number of stars. In the same way we find that for the XC1 metric the intersection is of 2,547 stars, translating into 82.3\% accuracy, and for XC2 it is 2,640 stars, leading to 81.6\% accuracy.

Because the three metrics have very different numerical scales, each metric will be standardised by removing the mean and scaling to unit variance before feeding it to the neural network. 
Therefore, the features derived from the unseen spectra to be vetted will also be standardised to the same mean and variance from the training set.

We are now interested in testing whether the combination of pairs of metrics allows for a better degree of separability than individually. We implement a simple linear Stochastic Gradient Descent classifier and fit it on the three pairs of metrics (AC-XC1, AC-XC2, XC1-XC2) plus their labels (reliable or unreliable), represented by 1s and 0s.  

Figure \ref{fig:metrics_ap} shows the linear decision function as a green line separating reliable from unreliable \dnu\ in the two-dimensional space of each pair of metrics. Setting a threshold as in Figure \ref{fig:metrics_ap}a for the pair AC-XC1 provides an accuracy in classification of 89.2\%, in (b) for the pair AC-XC2 the accuracy reaches 89.8\%.
Even when metrics XC1 and XC2 are based on the same characteristic of a star's power spectrum, there is indication that using both metrics is better than just selecting the best metric between them. 
The line separating the two populations
on the scatter plot from Figure \ref{fig:metrics_ap}c still improves the individual accuracy of metrics XC1 and XC2 from 82.3\% and 81.6\% respectively, to better than 84\% when combined.
Hence, AC, XC1, and XC2 %are complementary and 
constitute good inputs for the machine learning algorithm because individually they carry non-redundant information that strongly correlates with the visual vetting or target variable.

\section{Neural Network Vetting vs. Uncertainty Vetting}
\label{sec:vetting}
\begin{figure*}
\includegraphics[width=.7\textwidth]{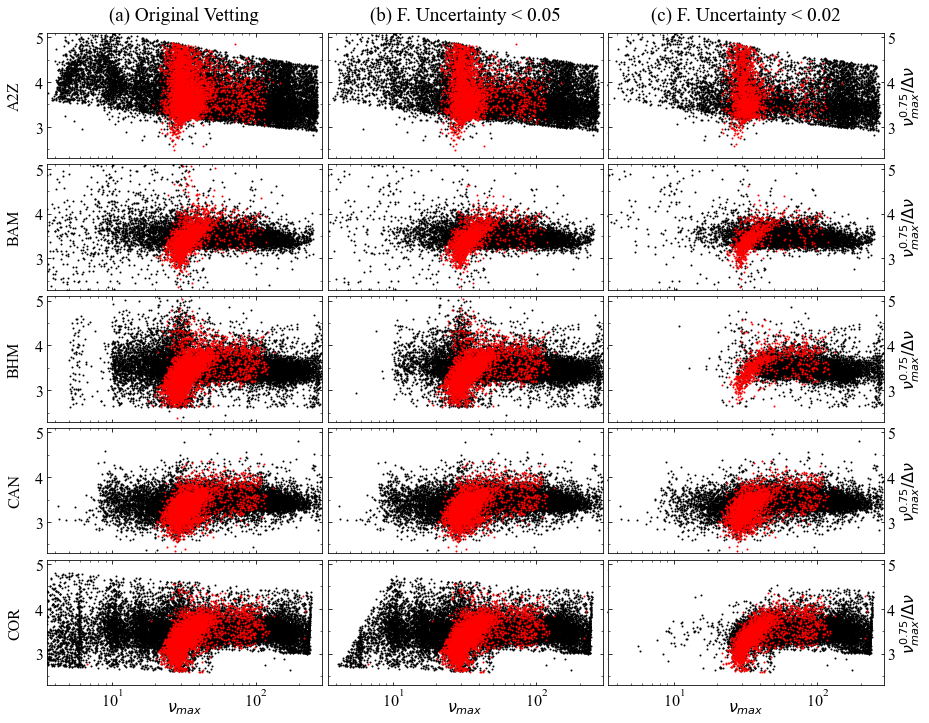}
    \caption{Vetting of \dnu{} using cuts based on fractional uncertainties up to 5\% and 2\%. RC stars appear in red and non-RC in black.}
    \label{fig:pipuncert}
\end{figure*}

We attempt to vet \dnu{} values from the five pipelines from Section \ref{sec:pipelines} (A2Z, BAM, BHM, CAN and COR) using only cuts in the \dnu\ fractional uncertainty for each pipeline's measurements. Figure \ref{fig:pipuncert}a, "Original Vetting" corresponds to the mass diagram of the same sample from the left column of Figure \ref{fig:pipes_grid}, making the same distinction between RC and non-RC stars using red and black. Columns b and c correspond to the resulting sample if we accept up to 5\% and 2\% of \dnu\ fractional uncertainty, respectively.

The performance of this uncertainty-vetting depends heavily on the pipeline and on the way their uncertainty is measured. For A2Z many of the clearly wrong values remain even after making the cut to fractional uncertainties lower than 2\%.
The CAN pipeline has very low fractional uncertainties across their entire sample, and neither of our cuts has a significant effect on it. For BAM, BHM and COR the cut to 2\% does help to bring out the characteristic "hook" formed by RC stars, however too many \dnu\ values are rejected in the lower \numax\ range.

\begin{figure*}
\includegraphics[width=.67\textwidth]{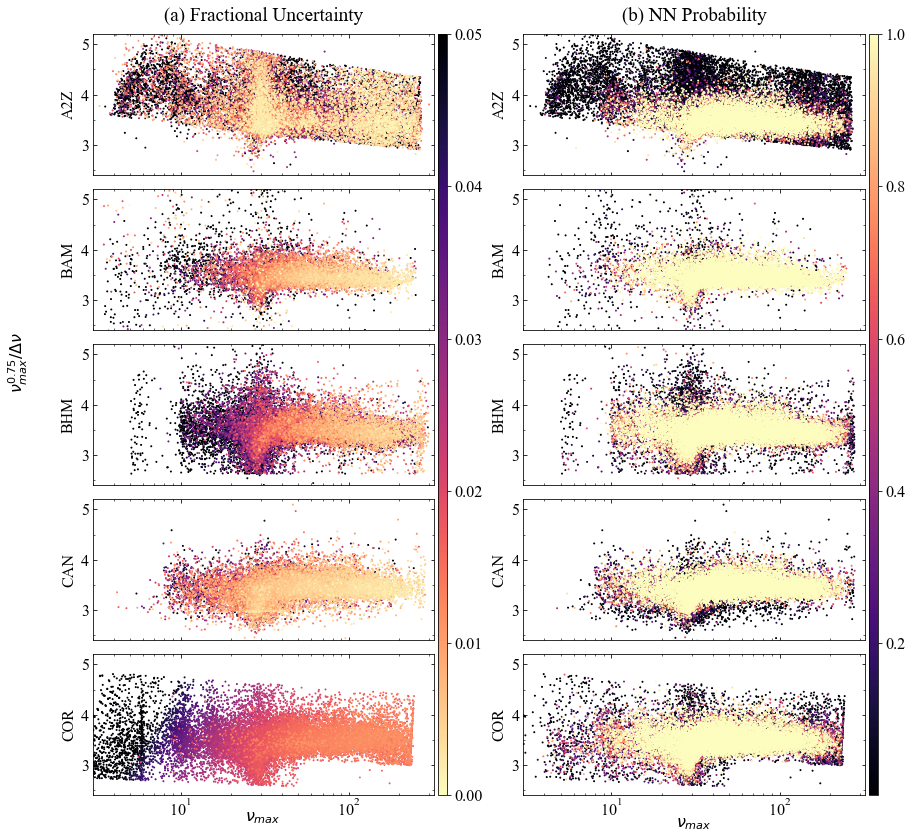}
    \caption{
    Diagrams showing $\nu_{\text{max}}^{0.75}/\Delta\nu$ as a function of \numax\ colour-coded based on (a) fractional \dnu\ uncertainties as derived by each pipeline, and (b) by the probabilities given by the neural network to the values provided by each pipeline.  }
    \label{fig:uncertainty}
\end{figure*}

Figure \ref{fig:uncertainty}a shows the original sample from each pipeline (same as Figure \ref{fig:pipuncert}a) colour-coded by \dnu\ fractional uncertainty, where every point with fractional uncertainty of 5\% and larger appears in black, and the lower fractional uncertainty points appear in yellow. Figure \ref{fig:uncertainty}b uses the same colour map to show the probabilities given by our neural network classifier. Note that the colour map is inverted because we look for low fractional uncertainty in (a) and for high probability in (b). We see that our classifier performs more consistently across the different pipelines, and decidedly removes those results that are clearly outliers and brings out in yellow the known shape of the mass proxy plot for every sample.

%%%%%%%%%%%%%%%%%%%%%%%%%%%%%%%%%%%%%%%%%%%%%%%%%%
%%%%%%%%%%%%%%%%%%%%%%%%%%%%%%%%%%%%%%%%%%%%%%%%%%
%%%%%%%%%%%%%%%%%%%%%%%%%%%%%%%%%%%%%%%%%%%%%%%%%%
%%%%%%%%%%%%%%%%%%%%%%%%%%%%%%%%%%%%%%%%%%%%%%%%%%
\section{Examples of apparent False Positives when comparing to ensemble-vetted sample}
\label{sec:missed}
\begin{figure*}
\includegraphics[width=\textwidth]{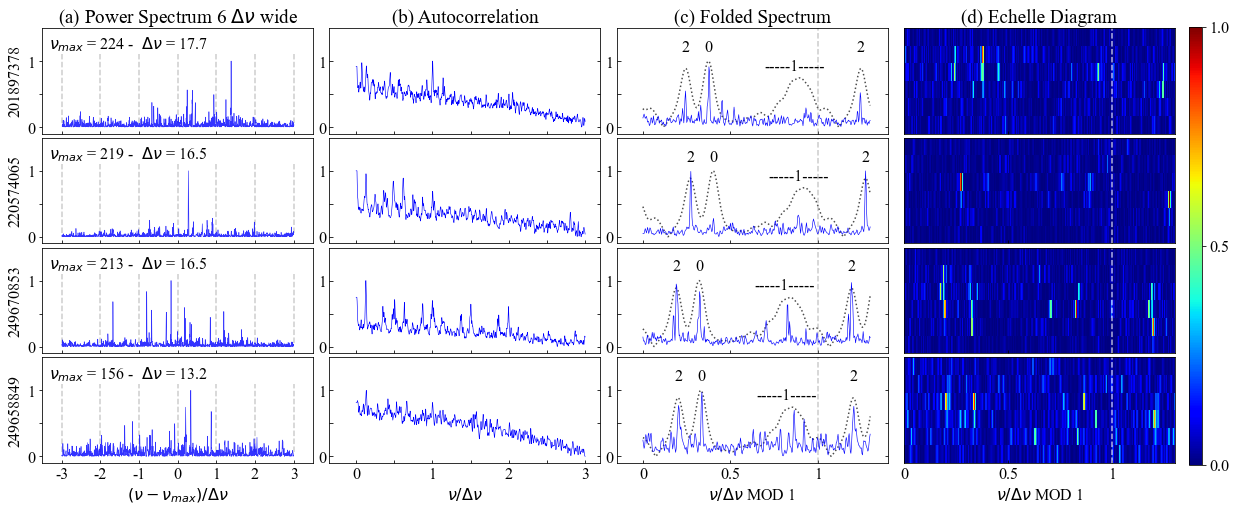}
    \caption{
    Examples of oscillating stars with \dnu\ considered good by the neural network, but missing from the ensemble-vetted sample. Values of \numax{} and \dnu{} are given in \muhz{} and all data have been scaled between 0 and 1.}
    \label{fig:sclump4}
\end{figure*}

In Figure \ref{fig:missing} we showed the mass diagram of the stars left out by the ensemble-vetting process, but accepted by our neural network. They are mostly RGB stars with \numax>100\muhz\ and most of them show clear oscillations. Here we show the diagnostic plots for four of them in Figure \ref{fig:sclump4}. These stars are proved to be True Positives, which is evident when looking at the folded spectra and especially the échelle diagrams.

\section{Examples of apparent False Negatives when comparing to ensemble-vetted sample}
\label{sec:error}

\begin{figure*}
\includegraphics[width=\textwidth]{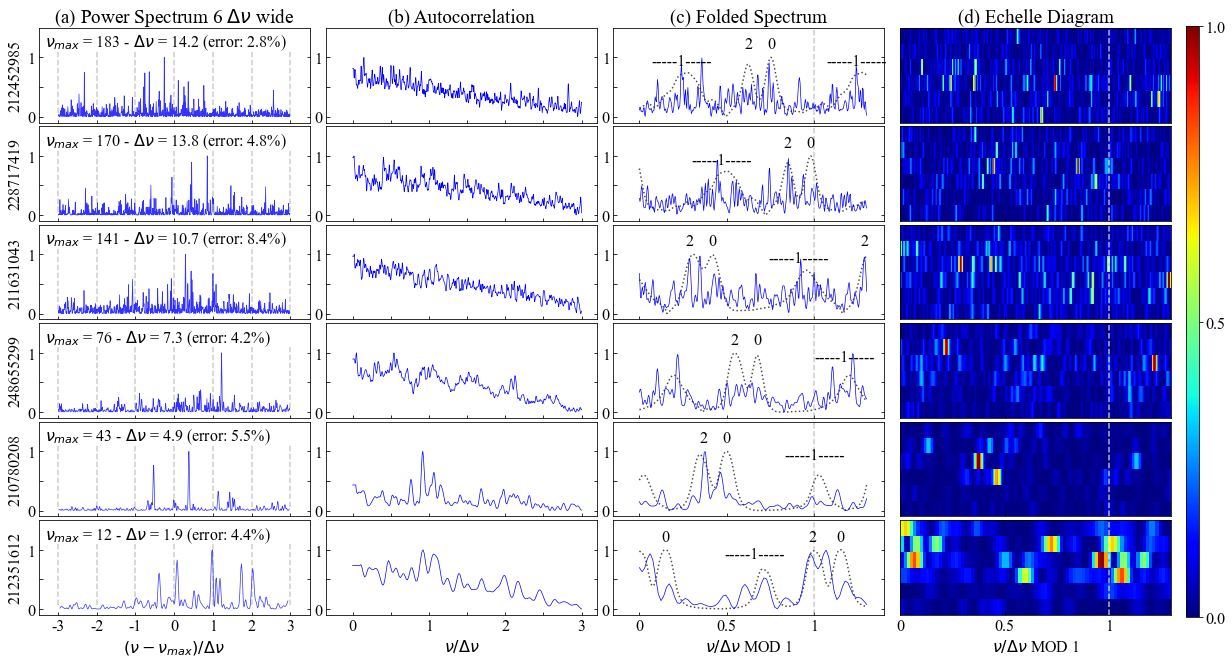}
    \caption{
    Diagnostic plots showing examples of \dnu\ from pipelines in Section \ref{sec:pipelines} rejected by the neural network. Values of \numax{} and \dnu{} annotated in column (a) are given in \muhz{} and the error with respect to our manually determined value of \dnu\ is in parenthesis.
    }
    \label{fig:rawerror}

\vspace{1cm}

\includegraphics[width=\textwidth]{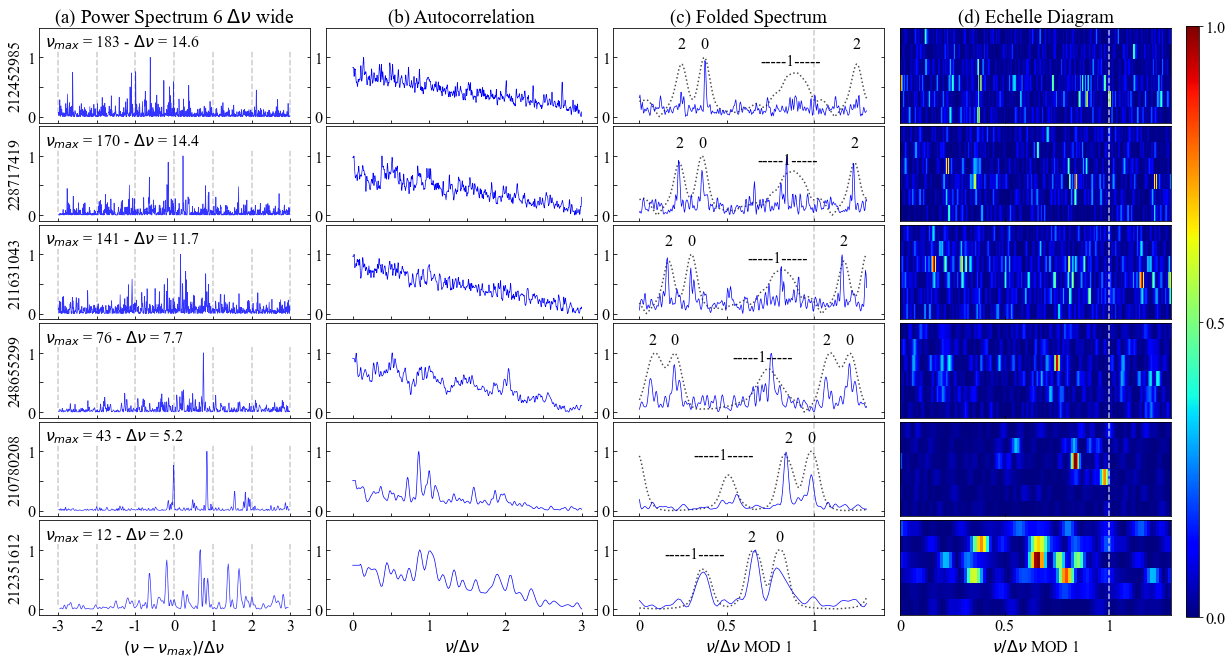}
    \caption{
    Diagnostic plots showing the same spectra from Figure \ref{fig:rawerror} but with \dnu\ visually determined as the value that puts the autocorrelation peaks closer to multiples of \dnu/2, makes the folded spectrum match the modelled template, and/or best aligns modes $l=2$ and $l=0$ in the échelle diagram. }
    \label{fig:corrected}
\end{figure*}

The analysis of \dnu\ values  
rejected by our classifier but accepted by the ensemble method revealed 
a higher number of suspected false negatives than expected from the network's measured performance, as shown in Table \ref{tab:mbypipe} and Figure \ref{fig:mbypipe2}.
A visual check of these rejected \dnu\ values showed that this higher number can be explained by the many cases where \dnu\ is offset by $\sim$3\% or more. 
These \dnu\ values were expected rejections due to the way our training sample and the network's features were constructed. 
For illustration, Figure \ref{fig:corrected} shows results of manually determined \dnu\ values found by visual inspection. This should be compared to Figure \ref{fig:rawerror}, which is based on raw pipeline \dnu\ values. For diagrams (c) and (d) in particular, the manual values show much better alignment.

\section{Tables}
\label{sec:tables}
We present in Table \ref{tab:syd_prob} our neural network vetted results after running it on the K2 sample with SYD parameters, including duplicates. In Table \ref{tab:pipes_prob} we present our neural network vetted results after running the network on K2 samples pre-vetted by each pipeline: A2Z, BAM, BHM, CAN and COR, using \numax\ and \dnu\ as derived by each pipeline and including duplicates. The threshold used to discriminate the good \dnu\ listed here was \textit{t=0.5}.  

\begin{table*}
    \caption{SYD results of K2 GAP sample after applying our neural network vetter. Column dnu\_prob indicates the probability assigned by our neural network. RC/RGB column indicates if the star was deemed to be RGB (0) or RC (1) by the machine learning method from \citet{2018MNRAS.476.3233H}. Columns numax\_sig and dnu\_sig indicate the uncertainty in the result by the SYD pipeline. Values in columns numax, numax\_sig, dnu and dnu\_sig are given in \muhz{}. This table contains 20,708 observations of 19,577 unique stars. Full table available as supplementary material.} 
	\centering
	\label{tab:syd_prob}
	\begin{tabular}{cccccccc} 
		\hline
		\multicolumn{8}{|c|}{Neural Network vetted results for values from SYD pipeline} \\
		\hline
		EPIC & campaign & numax & numax\_sig & dnu & dnu\_sig & dnu\_prob  & RC/RGB \\
		\hline
        201670988&	1&	4.539&	0.676&	1.001&	0.194&	0.540&	0\\
        201386006&	1&	5.088&	0.527&	1.031&	0.084&	0.751&	0\\
        201135864&	1&	5.117&	1.440&	1.098&	0.063&	0.624&	0\\
        201136194&	1&	5.504&	0.352&	1.081&	0.030&	0.910&	0\\
        201364846&	1&	5.868&	0.179&	1.244&	0.052&	0.829&	0\\
        \hline
	\end{tabular}
\end{table*}

\begin{table*}
    \caption{Neural Network vetted \dnu\ values for pipelines A2Z, BAM, BHM, CAN, and COR. Values in columns numax, numax\_sig, dnu, dnu\_sig are given in \muhz{}.
    Columns EV\_ensemble and EV indicate the evolutionary phase assigned to the star by the machine learning method from \citet{2018MNRAS.476.3233H} for ensemble-scaled \numax\ and \dnu\ values and for values of \numax\ and \dnu\ delivered by each pipeline, respectively.
    Column "dnu\_prob" indicates the probability assigned by our neural network. We have not removed stars with results from multiple campaigns. Full table available as supplementary material.}
	\centering
	\label{tab:pipes_prob}
	\begin{tabular}{cccccccccc} 
		\hline
		\multicolumn{10}{|c|}{Neural Network vetted results for \dnu\ values from Pipelines A2Z - BAM - BHM - CAN - COR} \\
		\hline
        Pipeline&	EPIC&	camp&	numax&	numax\_sig &	dnu &	dnu\_sig &	EV\_ensemble &	EV &	dnu\_prob \\
        \hline
        A2Z&	201703016&	1&	11.032&	0.698&	1.880&	0.071&	RGB/AGB&	RGB/AGB&	0.990\\
        A2Z&	201727507&	1&	11.990&	0.753&	2.020&	0.030&	RGB/AGB&	RGB/AGB&	0.656\\
        A2Z&	201627037&	1&	12.220&	0.786&	2.030&	0.325&	RGB/AGB&	RGB/AGB&	1.000\\
        A2Z&	201701753&	1&	12.248&	0.590&	1.730&	0.283&	RGB/AGB&	RGB/AGB&	1.000\\
        A2Z&	201553833&	1&	13.400&	2.039&	2.020&	0.009&	RGB/AGB&	RGB/AGB&	0.994\\
%        ...&	...&	...&	...&	...&	...&	...&	...&	...&	...\\
%        COR&	211764055&	18&	220.470&	3.780&	17.857&	0.225&	RGB&	NaN&	1.000\\
%        COR&	211741853&	18&	225.110&	3.550&	16.765&	0.207&	RGB&	NaN&	0.611\\
%        COR&	211742245&	18&	227.580&	4.150&	19.477&	0.258&	NaN&	NaN&	0.835\\
%        COR&	211940967&	18&	232.630&	3.820&	17.498&	0.259&	NaN&	NaN&	0.975\\
%        COR&	211732416&	18&	233.810&	3.590&	16.718&	0.227&	RGB&	NaN&	0.762\\
        \hline
	\end{tabular}
\end{table*}
% Don't change these lines
\bsp	% typesetting comment
\label{lastpage}
\end{document}